\documentclass[prl,aps,reprint,floatfix,showpacs,twocolumn,superscriptaddress]{revtex4-1}
\usepackage{amssymb,amsfonts,amsmath}
\usepackage{graphics}
\usepackage[dvips]{graphicx}
\usepackage{color}
\usepackage{subfigure}
\usepackage{mathtools}
\usepackage[normalem]{ulem}
\usepackage{dcolumn}
\usepackage{multirow}
\usepackage{booktabs}
\usepackage{bm}
\usepackage[linktocpage,colorlinks=true,linkcolor=blue,citecolor=blue]{hyperref}

\newcommand{\mf}[1]{\ensuremath{\mathbf{#1}}}
\newcommand{\la}{\langle}
\newcommand{\ra}{\rangle}

\newcommand{\mrm}[1]{\ensuremath{\mathrm{#1}}}

\newcommand{\ovb}[2]{\ensuremath{\overbracket[0.5pt][1.5pt]{\,\mf{\widehat{#1}}^{(#2)}}}}
\newcommand{\unv}[1]{\ensuremath{\hat{\mf{#1}}}}

\begin{document}

% ======================== TITLE ========================
% \title{Irreducible representations of oscillatory and swirling flows around microswimmers}
\title{Irreducible representations of oscillatory and swirling flows in active soft matter}

% ======================== AUTHORS, AFFILIATIONS ========================
\author{Somdeb Ghose}
\affiliation{The Institute of Mathematical Sciences, CIT Campus, Chennai 600113, India}
\author{R. Adhikari}
\affiliation{The Institute of Mathematical Sciences, CIT Campus, Chennai 600113, India}

% ======================== DATE ========================
% \date{April 1, 2012}
\date{\today}

% ======================== ABSTRACT ========================
\begin{abstract}
Recent experiments imaging fluid flow around swimming microorganisms have revealed complex time-dependent velocity fields that differ qualitatively from the stresslet flow commonly employed in theoretical descriptions of active matter. Here we obtain the most general flow around a finite sized active particle by expanding the surface stress in irreducible Cartesian tensors. This expansion, whose first term is the stresslet, must include, respectively, third-rank polar and axial tensors to minimally capture crucial features of the active oscillatory flow around translating Chlamydomonas and the active swirling flow around rotating Volvox. The representation provides explicit expressions for the irreducible symmetric, antisymmetric and isotropic parts of the continuum active stress. Antisymmetric active stresses do not conserve orbital angular momentum and our work thus shows that spin angular momentum is necessary to restore angular momentum conservation in continuum hydrodynamic descriptions of active soft matter.
\end{abstract}

% \pacs{}

\maketitle

% ===============================================================================================
% INTRO
% ===============================================================================================
The collective dynamics of microscopic particles that swim in viscous fluids by converting chemical energy to mechanical work is a topic of current interest in non-equilibrium statistical mechanics \cite{pedley1992, *cisneros2007, *lauga2009, *ramaswamy2010, *cates2011, *koch2011, *lauga2012, *marchetti2013}. Biological and biomimetic examples of such ``active'' particles include molecular motors \cite{nedelec1997}, active nanobeads \cite{paxton2004, *vicario2005, *catchmark2005, *fournier2005, *dreyfus2005}, ATP driven biomimetic systems \cite{sanchez2011, *sanchez2012}, light-activated colloidal surfers \cite{palacci2013} and swimming microorganisms. 
Momentum conservation and the lack of inertia at the microscopic scale imply that the fluid flow around such particles must be both force-free and torque-free, thus constraining it to decay no slower than the inverse square of the distance from the particle. Thus, at distances large compared to the particle size the dominant contribution to the flow is from the dipolar stresslet singularity \cite{batchelor1970}. Continuum theories, applicable at scales much larger than the particle size, employ the stresslet flow to obtain the long-wavelength, long-time features of the collective dynamics of active suspensions \cite{simha2002a, *saintillan2008a}.

The flow around one class of active particles, swimming microorganisms, has been resolved in unprecedented spatial and temporal detail in recent experiments \cite{drescher2009, *guasto2010, *drescher2010}. These reveal near field features that cannot be captured by the standard purely stresslet description \cite{batchelor1970, blake1971c, simha2002a, saintillan2008a}. The complex flow around Chlamydomonas has easily identifiable qualitative features like stagnation points and strong lateral circulations that vary periodically with time. Both Chlamydomonas and Volvox rotate about their axis while swimming \cite{guasto2010, drescher2009} and thus generate swirling flows. 
Further, the experimentally measured value of the power dissipated in swimming cannot be computed from the stresslet singularity, which dissipates an infinite amount of power. 

The above experiments point to the need of a time-dependent description of active microswimming that accounts for both axisymmetric and swirling flows, and instead of constructing the flow from singularities, obtains it from the governing equations by satisfying the fluid boundary conditions that prevail on the surface of a finite-sized swimmer.  These boundary conditions, which may prescribe stresses or velocities, must be able to produce both particle translations and particle rotations. Stresslet flow can do neither and thus it is imperative to identify the minimal set of independent stress modes that can produce a general rigid body motion of the particle. 

With these motivations we present, in this Letter, the most general representation of Stokes flow around a finite-sized active particle as an expansion in irreducible Cartesian multipoles of the surface stress. The orthogonality and completeness of the tensorial multipoles provide simple relations between the stresses and velocities that allow us to identify the multipoles necessary and sufficient for active translation and rotation.  Knowing the rigid body motion we are thus able to reconstruct, using only a few irreducible multipoles, the complex time-dependent flows observed in experiment.  The power dissipation and swimming efficiency obtained in terms of these multipoles are in good agreement with experiment. 

The irreducible tensor expansion provides a particle level description for studying the collective microhydrodynamics of active soft matter \cite{pedley1992, *cisneros2007, *lauga2009, *ramaswamy2010, *cates2011, *koch2011, *lauga2012, *marchetti2013}. We illustrate this by formally deriving a constitutive equation for the active continuum stress in terms of densities of the irreducible multipoles. Previously, symmetry arguments were used to derive the first term of this constitutive equation \cite{simha2002a}. Our formal derivation reveals the generic presence of antisymmetric stresses in the constitutive equation, which lead to counterintuitive effects such as the separate global conservation of orbital and spin angular momentum and the generation of macroscopic flows in suspensions of spinning particles. Our results show that extant continuum descriptions of active soft matter based on second-rank symmetric tensor order parameters are incomplete, and the formalism developed in this work provide the basis for the most general description in terms of higher rank order parameters.  

% ====================================================
% MODEL
% ====================================================
\emph{Irreducible representations of Stokes flow:} Creeping flow around a particle obeys the Stokes equation, $\nabla \cdot \bm{\sigma} = - \nabla p + \eta \nabla^2 \mf{v} = 0$, $\nabla \cdot \mf{v} = 0$, where $\mf{v}$ is the flow within the volume $V$, $\bm{\sigma}$ is the stress, $p$ is the pressure and $\eta$ is the viscosity. Chemomechanical activity can regulate either the velocity $\mf{v}^S$ or the stress  $\bm{\sigma}^S$ on the surface $S$ of the particle, which requires Dirichlet or Neumann boundary conditions, respectively. In either case, the flow in the bulk can be expressed as an integral over the boundary $S$, where a single layer density $\mf{q}(\mf{r})$ is convolved with the dyadic Green's function $\mf{G}(\mf{r}) = (\mathbb{I} + \unv{r} \, \unv{r}) / |\mf{r}|$ \cite{ladyzhenskaya1969, *pozrikidis1992, *kim2005}
\begin{align} \label{eq:BI_representation_and_equation}
  \int_{S'}
  \mf{G}(\mf{r} - \mf{r}') \cdot \mf{q}(\mf{r}') \, \mrm{d} S'
  &=
  - 8 \pi \eta \, 
  \begin{cases}
    \mf{v}(\mf{r}), \,\, \mf{r} \in V
    \\
    \mf{v}^S(\mf{r}), \,\, \mf{r} \in S,
  \end{cases}
\end{align}
where $\mf{r}$ is the field point in the bulk $V$ and $\mf{r}'$ is the source point on the surface $S$. Eq.\ (\ref{eq:BI_representation_and_equation}), $\mf{r} \in V$, provides a complete solution for the Neumann problem with known single layer density. For the Dirichlet problem, Eq.\ (\ref{eq:BI_representation_and_equation}), $\mf{r} \in S$, must be solved to obtain the unknown single layer density in terms of the prescribed boundary velocity $\mf{v}^S$.

We expand the single layer stress density on a sphere of radius $a$, which may represent the physical boundary of the particle or a spherical surface enclosing the particle, in terms of irreducible Cartesian tensors $\ovb{r}{p}$, \cite{coope1965, *jerphagnon1970, *jerphagnon1978, *mazur1982, *ladd1988}
\begin{align} \label{eq:traction_jump_irred_expansion}
  \mf{q}(\mf{r}) 
  &= 
  \sum_{p=0}^{\infty}
  \frac{( 2p + 1 ) !!}{4 \pi a^2} \,
  \, \ovb{r}{p} \odot \, \mf{Q}^{(p+1)}, \, \mf{r} \in S
\end{align}
where the multipole moments $\displaystyle {Q}^{(p+1)}_{i\alpha_1 \ldots \alpha_p}$, symmetric and traceless in the last $p$ indices, are given by $p! \, \mf{Q}^{(p+1)} = \int \mf{q}(\mf{r}) \ovb{r}{p} \, \mrm{d} S$. Here $\odot$ indicates a $p$-fold contraction between a $p$-th rank tensor and another of higher rank, contracting the last index of the first tensor with the first index of the latter till $p$ indices are contracted, such that $\ovb{r}{p} \odot \, \mf{Q}^{(p+1)} = \overbracket[0.7pt][3.0pt]{\widehat{r}_{\alpha_1 \alpha_2 \ldots \alpha_{p-1} \alpha_p}} {Q}^{(p+1)}_{\alpha_p \alpha_{p-1} \ldots \alpha_2 \alpha_1 \, i}$. 
The $p$th rank tensor $\ovb{r}{p}$ is symmetric and traceless in every pair of its $p$ indices and obeys the orthogonality relation $(2p+1)!! \, \la \ovb{r}{p} \ovb{r}{q} \ra = 
  p! \, \delta_{p,q} \, \bm{\Delta}^{(p,p)}$.
% 
% \begin{align} \label{eq:irred_orthogonality}
%   (2p+1)!! \, \Big\la \ovb{r}{p} \ovb{r}{q} \Big\ra 
%   &= 
%   p! \, \delta_{p,q} \, \bm{\Delta}^{(p,p)}.
% \end{align}
% 
The surface average $\la \ldots \ra = (1/4 \pi a^2) \int \mrm{d} S$, while the rank $2p$ tensor $\bm{\Delta}^{(p,p)}$ projects any $p$-th rank tensor $\mf{\widehat{r}}^{(p)}$ to its irreducible form $\ovb{r}{p}$ \cite{hess1980, mazur1982}. 

Substituting Eq.\ (\ref{eq:traction_jump_irred_expansion}) into the boundary integral Eq.\ (\ref{eq:BI_representation_and_equation}), and following the method detailed in the supplemental material \cite{SI}, we obtain the solution of the Stokes equation
as $\mf{v}(\mf{r}) = \sum_{p=0}^{\infty} (-1)^{p+1} \, \mf{v}_p (\mf{r})$, where 
\begin{align} \label{eq:stokes_flow_for_single_sphere}
  8 \pi \eta \, \mf{v}_p(\mf{r})
  &=
  a^p \, \mf{Q}^{(p+1)} \odot 
  \left( 1 + \frac{a^2 }{4p+6} \nabla^2 \right) 
  \bm{\nabla}^{(p)} \mf{G}(\mf{r}).
\end{align}
The flow $\mf{v}_p$ at any order $p$ has contributions which decay as $r^{-p}$ and $r^{-(p+2)}$. Thus the stress multipole expansion automatically generates the Fax\'{e}n corrections $a^2 \nabla^2 \mf{G}(\mf{r}) /(4p+6) $ that must be manually reconstructed when expanding in the velocity multipoles of Lamb's general solution \cite{lamb1916}.

The reducible surface stress multipoles of rank $p$ can be expressed as a direct sum of  their irreducible parts, $\mf{Q}^{(p)} = \oplus_{j} \mf{Q}^{(p; \, j)}$ \cite{SI}, indexed by their weights $j\leq p$. The constraints imposed by incompressibility, biharmonicity and spherical symmetry imply that, at each order, only the three highest irreducible parts contribute. Here we focus on the minimal set of multipoles required to produce active translations and rotations.
The decompositions we require are \cite{coope1965, *jerphagnon1970, *jerphagnon1978, andrews1982, SI}
$\mf{Q}^{(1)} = \mf{F}$, 
$a \, \mf{Q}^{(2)} 
=  \mf{S} - \frac12 \bm{\epsilon} \cdot \mf{T} $, 
$a^2 \, \mf{Q}^{(3)}_{i \alpha \beta}
= \bm{\Gamma}_{i \alpha \beta}
+ \frac13 
\big\{ 
\bm{\epsilon} \cdot \bm{\Psi} + \left( \bm{\epsilon} \cdot \bm{\Psi} \right)^{\mrm{T}} 
\big\}_{i \alpha \beta}
+ \frac{1}{10} \left(-2 \mf{d}_i \mathbb{I}_{\alpha \beta} 
+ 3 \mf{d}_{\alpha} \mathbb{I}_{\beta i} 
+ 3 \mf{d}_{\beta} \mathbb{I}_{i \alpha} \right) $, 
$a^3 \, \mf{Q}^{(4; \, 3)} 
= - \frac{3}{4} \overbrace{\bm{\epsilon} \cdot \bm{\Lambda}} $
where $\overbrace{\ldots}$ denotes complete symmetrization and $\bm{\epsilon}$ is the rank-3 antisymmetric Levi-Civita tensor. The force $\mf{F}$, the torque $\mf{T}$, stresslet $\mf{S}$ and the potential dipole $\mf{d}$ are familiar irreducible multipoles. The new irreducible multipoles introduced here are the second rank pseudodeviatoric torque dipole $\bm{\Psi}$ or the ``vortlet'', the third rank septorial stresslet dipole $\bm{\Gamma}$ or the ``septlet'', and the third rank pseudoseptorial multipole $\bm{\Lambda}$ or the ``spinlet''. Using these decompositions and Eq.\ (\ref{eq:stokes_flow_for_single_sphere}), force-free torque-free flows decaying no faster than $r^{-5}$ are expressed as
\begin{align} \label{eq:flows_due_to_irred_multipole_moments}
  &8 \pi \eta \, \mf{v}^{\mrm{act}} (\mf{r})
  \! = \!
  \left( \! 1 + \frac{a^2}{10} \nabla^2 \! \right)
  \! \bm{\nabla} \mf{G} \! \odot \! \mf{S}
% \\
% 
  + \frac15 \nabla^2 \mf{G} \! \cdot \! \mf{d}
\nonumber \\
&
  + \frac23 \!
  \left( \! \bm{\Psi} \cdot \bm{\nabla} \! \right)
  \! \cdot \!
  \left( \bm{\nabla} \! \times \! \mf{G} \right)
% \nonumber \\
% & 
  - 
  \left(  1 + \frac{a^2}{14} \nabla^2  \right)
  \bm{\nabla} \bm{\nabla} \mf{G} \odot \bm{\Gamma}
\nonumber \\
&
  - \frac34
  \left( \bm{\Lambda} : \bm{\nabla} \bm{\nabla} \right)
  \cdot \left( \bm{\nabla} \times \mf{G} \right).
\end{align}
The total number of independent coefficients is $5+3+5+7+7 = 27$. 
The stresslet $\mf{S}$ completely characterizes active flows decaying as $r^{-2}$. The potential dipole $\mf{d}$, the vortlet $\bm{\Psi}$ and the septlet $\bm{\Gamma}$ together completely characterize flows decaying as  $r^{-3}$. The spinlet $\bm{\Lambda}$ produces a flow decaying as $r^{-4}$. The vortlet and the spinlet produce swirling flows which have not been considered before. 
These flows are plotted in \cite{SI}.

% ====================================================
% FIGURE # : GOLLUB
% ====================================================
\begin{figure*}[tbp]
  \begin{center}
    \subfigure{ 
    \includegraphics[width=0.27\textwidth, height=0.16\textheight]
    {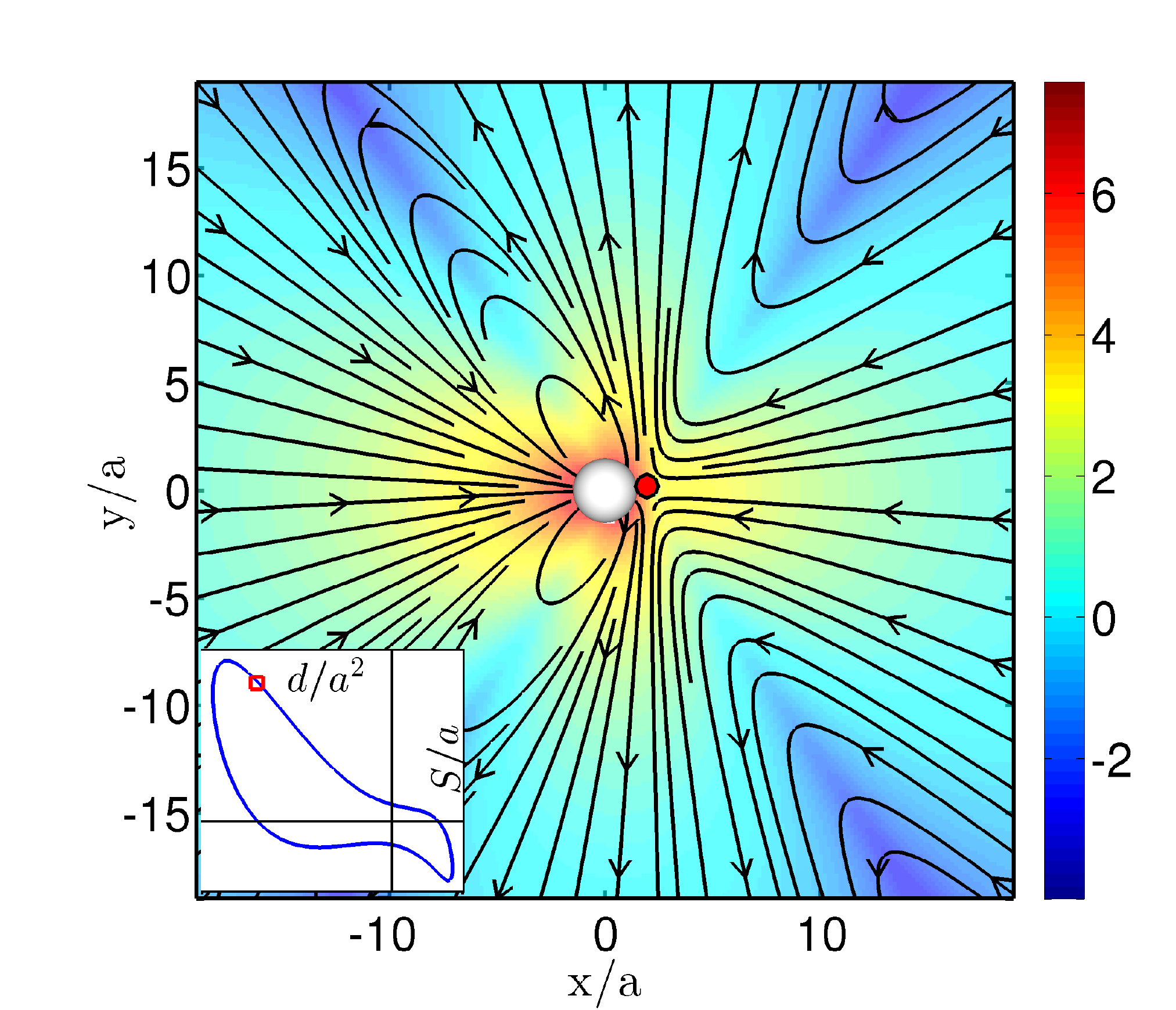}
%    {gollub2010PRL_3a_g} 
    \label{fig:gollub2010PRL_a} 
    } 
    \subfigure{ 
    \includegraphics[width=0.27\textwidth, height=0.16\textheight]
    {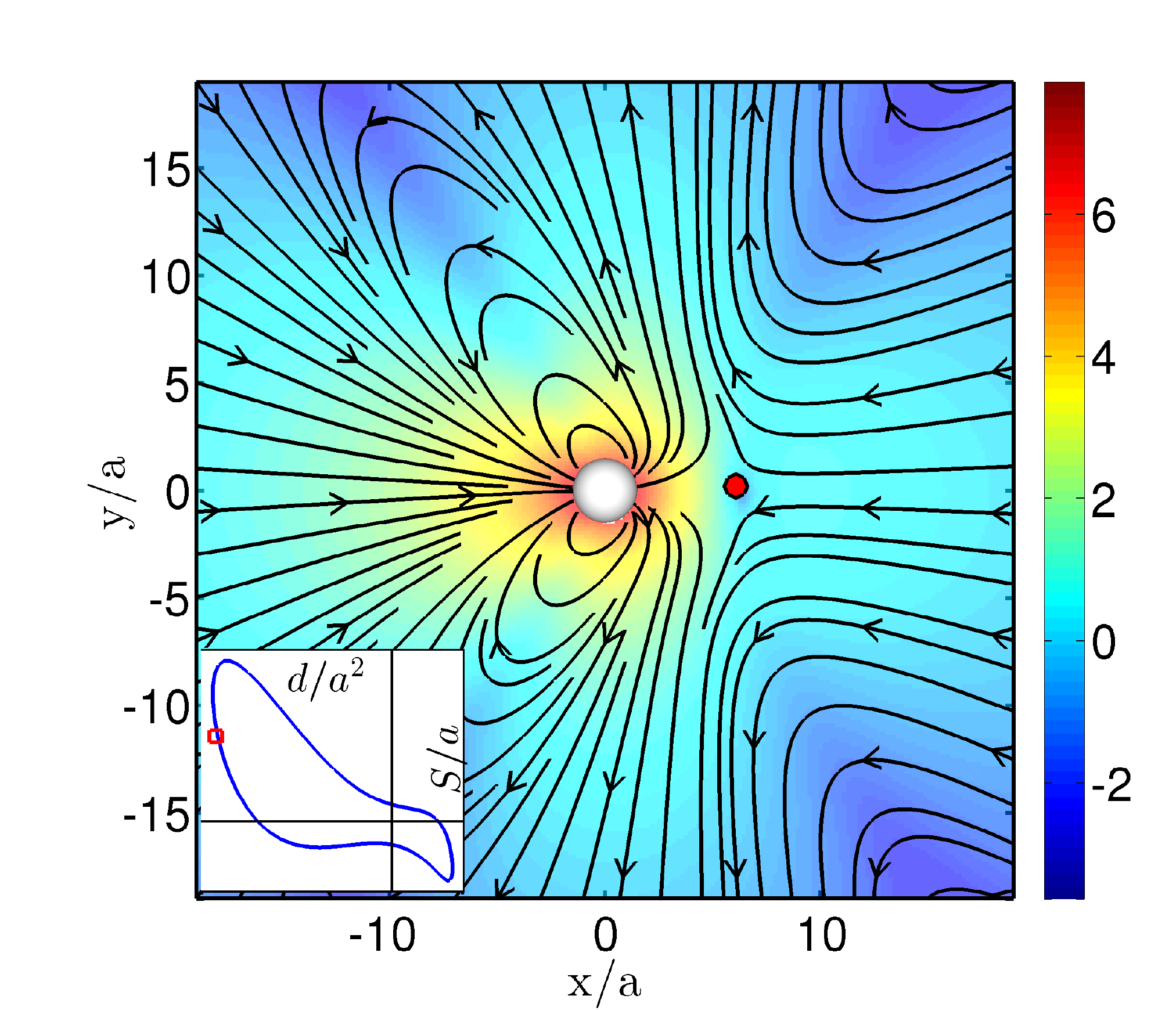} 
 %    {gollub2010PRL_3b_g} 
    \label{fig:gollub2010PRL_b} 
    } 
    \subfigure{ 
    \includegraphics[width=0.27\textwidth, height=0.16\textheight]
    {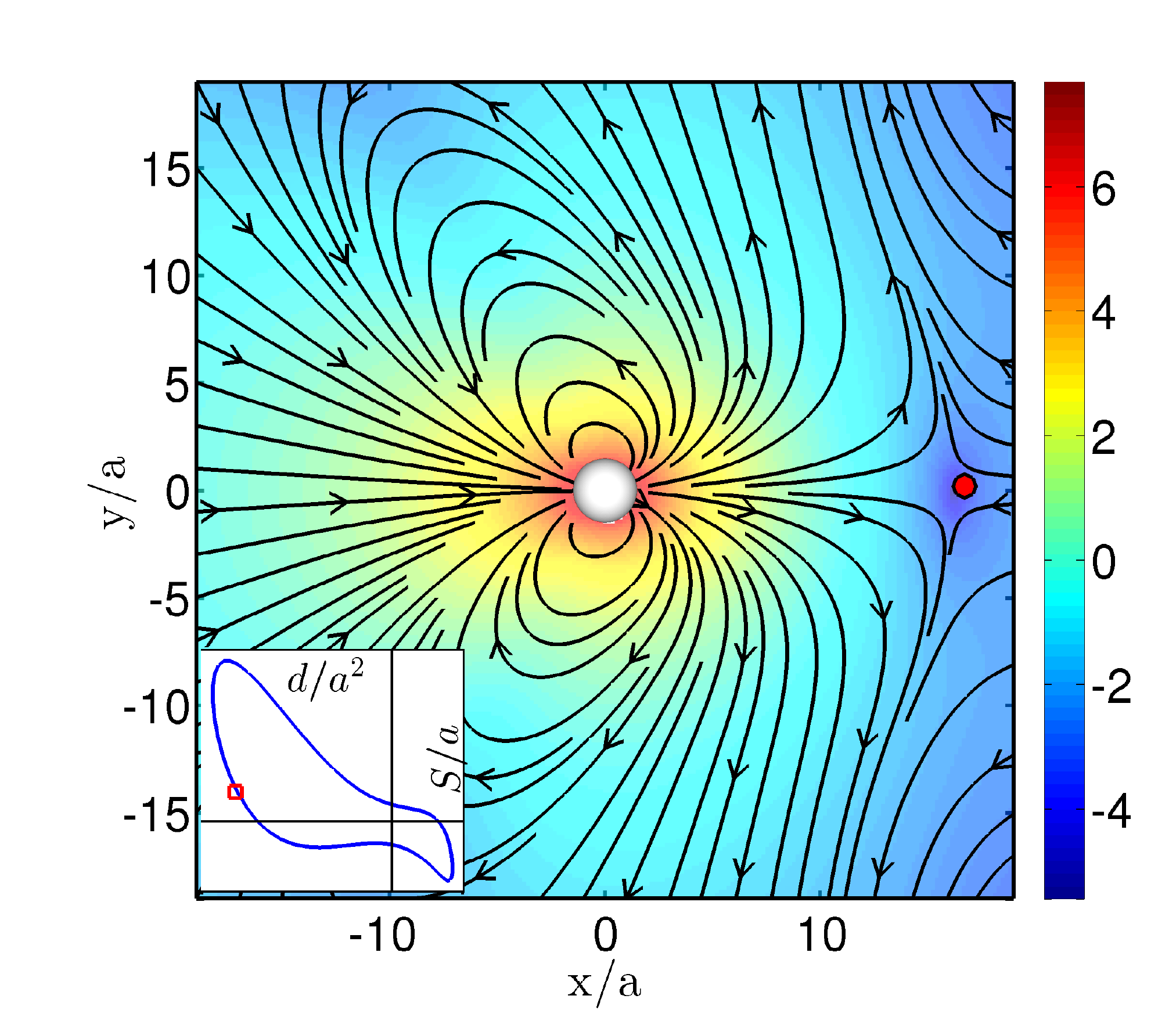} 
 %    {gollub2010PRL_3c_g} 
    \label{fig:gollub2010PRL_c} 
    } \\
    \subfigure{ 
    \includegraphics[width=0.27\textwidth, height=0.16\textheight]
    {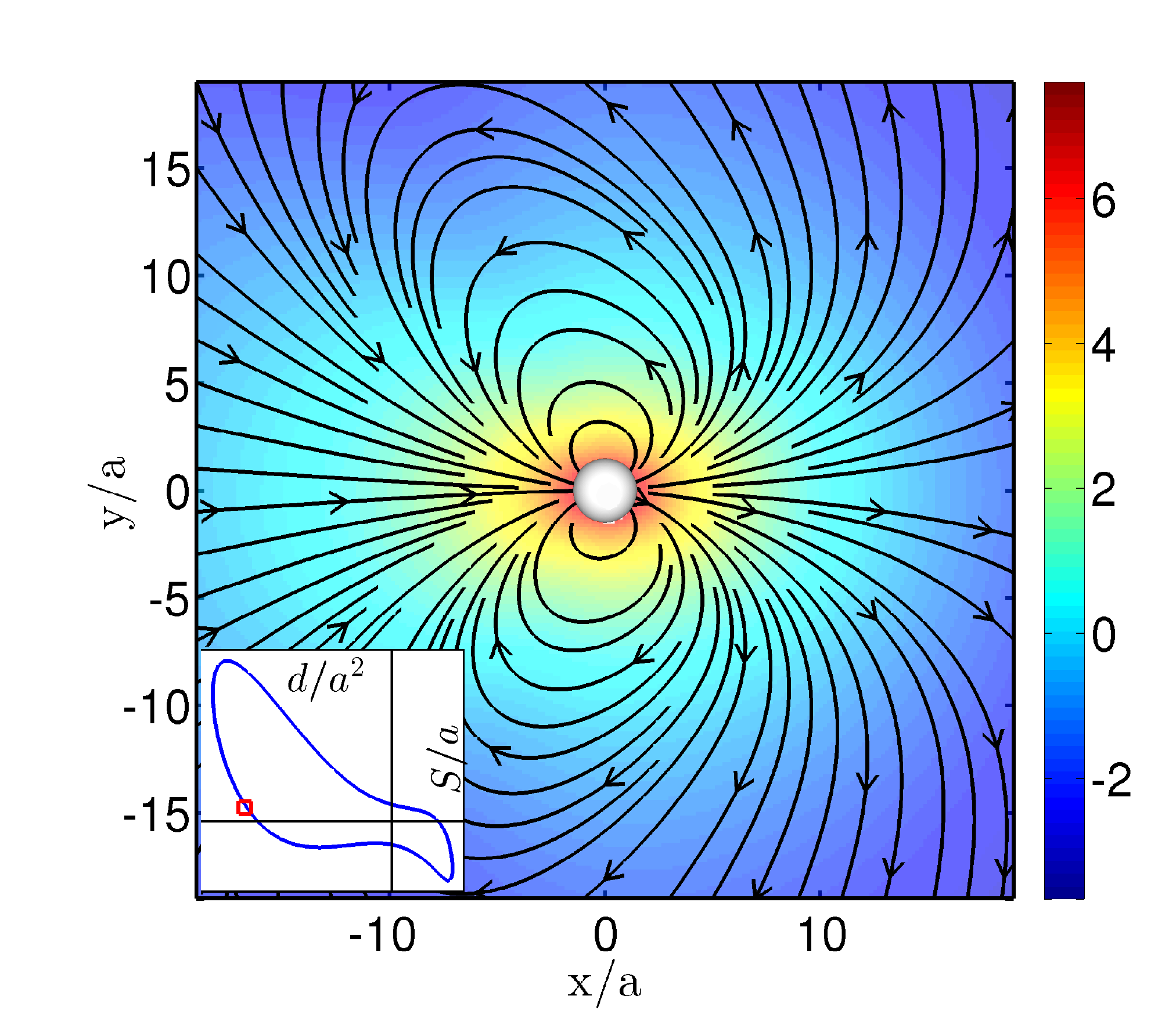} 
 %    {gollub2010PRL_3d_g} 
    \label{fig:gollub2010PRL_d} 
    } 
    \subfigure{ 
    \includegraphics[width=0.27\textwidth, height=0.16\textheight]
    {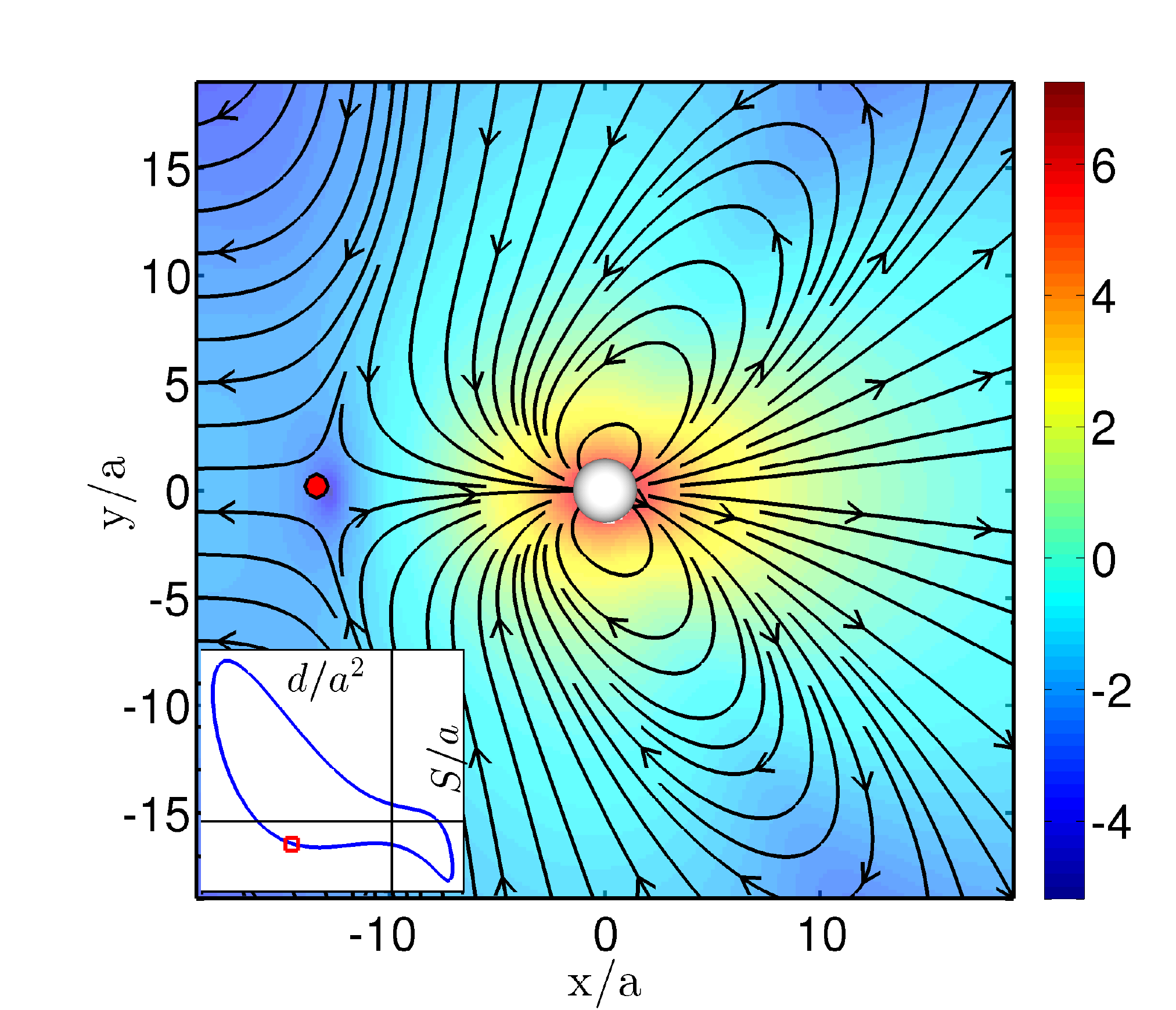}
 %    {gollub2010PRL_3e_g} 
    \label{fig:gollub2010PRL_e} 
    } 
    \subfigure{ 
    \includegraphics[width=0.27\textwidth, height=0.16\textheight]
    {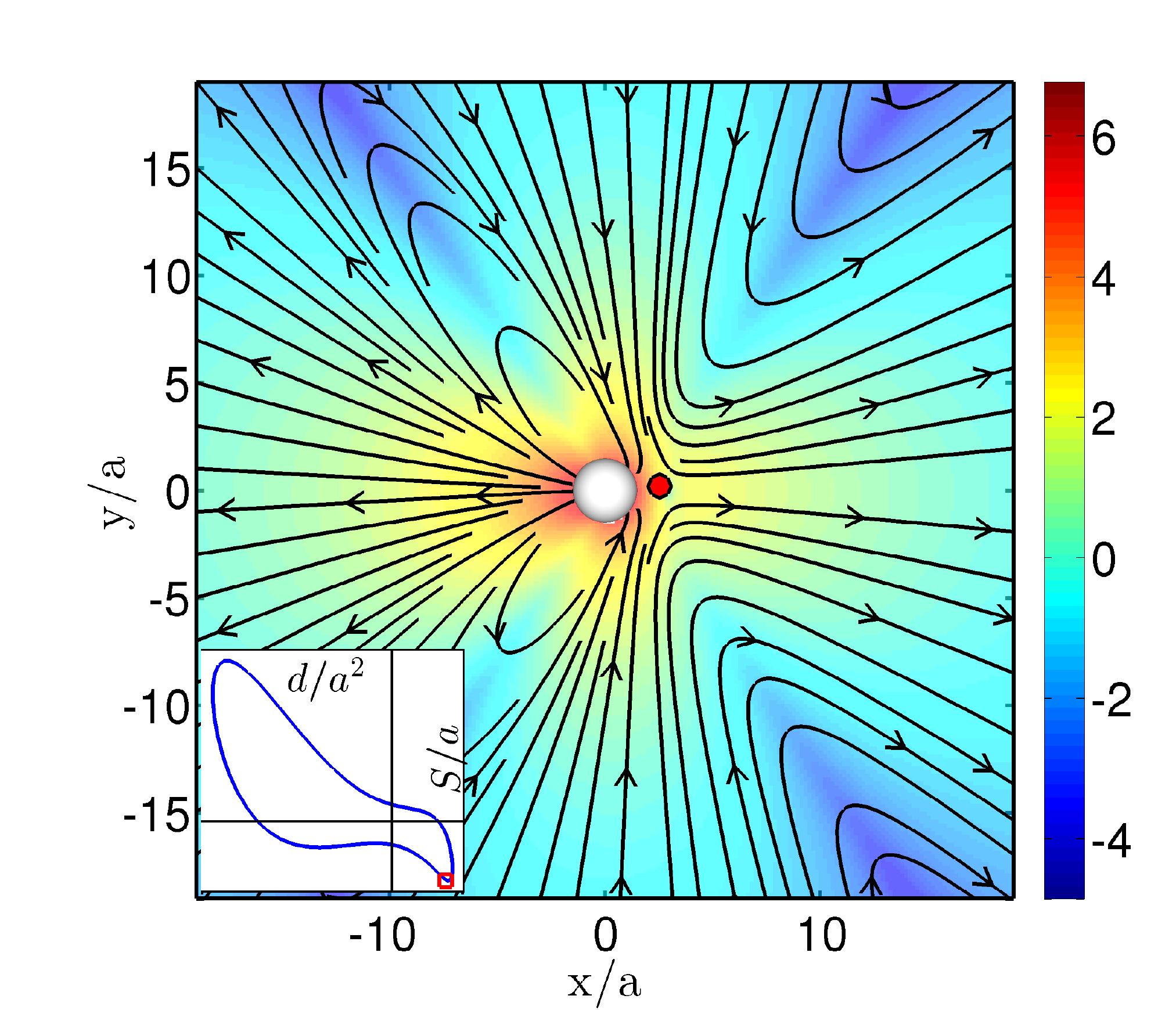}
 %    {gollub2010PRL_3f_g} 
    \label{fig:gollub2010PRL_f} 
    }
  \caption{ (Color) Cross-sections of complex time-dependent flows around a spherical active particle generated by a simple linear combination of $\mf{S}(t)$, $\mf{d}(t)$ and $\bm{\Gamma}(t)$ that capture essential features of flow fields around swimming Chlamydomonas \cite{guasto2010}. Streamlines show the direction while the background color represents the natural logarithm of the strength of the flow. The red dot indicates the approximate position of the stagnation point. The inset shows the overall (blue solid line) and instantaneous (red dot) variation of $d_0(t)$ against $S_0(t)$.
  \label{fig:gollub_chlamy}}
  \end{center}
\end{figure*}
% ====================================================

% ====================================================
% SECTION : Active particle motion
% ====================================================
\emph{Active particle motion :}
The active translations  $\mf{V}$ and rotations $\bm{\Omega}$ of the \emph{particle} can be obtained from the linear relation between the velocity and stress multipoles at the boundary, where $\mf{v}^S(\mf{r}) = \mf{V} + a \bm{\Omega} \times \widehat{\mf{r}} + \mf{v}^{\mrm{a}}(\mf{r})$ and  $\mf{v}^{\mrm{a}}$ is an activity induced surface velocity. 
Expanding the surface velocity $\mf{v}^S(\mf{r})$ in the same irreducible basis $\ovb{r}{p}$ and using Eq.\ (\ref{eq:BI_representation_and_equation}), $\mf{r} \in S$,
the linear relationships can be expressed explicitly as \cite{SI}
\begin{subequations} \label{eq:single_sphere_Laddyzhenskaya_relations}
  \begin{align}
  \label{eq:single_sphere_Laddyzhenskaya_Force}
    &
    \mf{F} 
%     \mf{0}
    =
    - 6 \pi \eta a \, 
    \Big(
    \mf{V} +
    \big\langle \mf{v}^{\mrm{a}} \big\rangle 
    \Big)
\\
  \label{eq:single_sphere_Laddyzhenskaya_Torque}
    &
    \mf{T} 
%     \mf{0}
    =
    - 8 \pi \eta a^2 \,     
    \Big(
    a \bm{\Omega} -
    \frac32 \Big\langle \mf{v}^{\mrm{a}} \times \widehat{\mf{r}} \Big\rangle 
    \Big)
\\
  \label{eq:single_sphere_Laddyzhenskaya_Stresslet}
    &\mf{S} 
    =
    - 10 \pi \eta a^2 \, 
    \big\langle 
    \mf{v}^{\mrm{a}} \, \widehat{\mf{r}} 
    + \left( \mf{v}^{\mrm{a}} \, \widehat{\mf{r}} \right)^{\mrm{T}}
    \big\rangle 
\\
  \label{eq:single_sphere_Laddyzhenskaya_PotentialDipole}
    &\mf{d} 
    = 
    - 30 \pi \eta a^3 
    \Big\langle 
    \, (\mf{v}^{\mrm{a}} \cdot \widehat{\mf{r}}) \, \widehat{\mf{r}} - \frac13 \mf{v}^{\mrm{a}}
    \Big\rangle
\\
  \label{eq:single_sphere_Laddyzhenskaya_TorqueDipole}
    &\bm{\Psi} 
    = 
    5 \pi \eta a^3 
    \Big\langle
    ( \mf{v}^{\mrm{a}} \times \widehat{\mf{r}} ) \, \widehat{\mf{r}}
    + \left\{ ( \mf{v}^{\mrm{a}} \times \widehat{\mf{r}} ) \, \widehat{\mf{r}} \right\}^{\mrm{T}}
    \Big\rangle
\\
  \label{eq:single_sphere_Laddyzhenskaya_StressletDipole}
    &\bm{\Gamma}
    =
    - \frac{35}{2} \pi \eta a^3
    \Big \langle
    \overbrace{\, \mf{v}^{\mrm{a}} \, \widehat{\mf{r}} \widehat{\mf{r}} \,}
    - \frac25 (\mf{v}^{\mrm{a}}\cdot \widehat{\mf{r}}) 
    \overbrace{\widehat{\mf{r}} \, \mathbb{I} }
    - \frac15 \overbrace{\mf{v}^{\mrm{a}}\, \mathbb{I} }
    \Big \rangle
\\
  \label{eq:single_sphere_Laddyzhenskaya_Spinlet}
   & \bm{\Lambda}
   =
   14 \pi \eta a^4
   \Big \langle
   \overbrace{\, \big(\mf{v^{\mrm{a}}} \times \widehat{\mf{r}} \big) \, \widehat{\mf{r}} \, \widehat{\mf{r}} \,}
   - \frac35
   \overbrace{\big(\mf{v^{\mrm{a}}} \times \widehat{\mf{r}} \big) \, \mathbb{I} \, }
   \Big \rangle.
 \end{align}
\end{subequations}
The first two relations above show that a force-free torque-free particle acquires translation and rotational motion only if the surface averages of the $\mf{v}^{\mrm{a}}$ and $\mf{v}^{\mrm{a}} \times \unv{r}$ are non-zero \cite{brenner1963, *anderson1991, *stone1996}. 
A uniaxial version of Eq.\ (\ref{eq:single_sphere_Laddyzhenskaya_Stresslet}), applicable only to axisymmetric flows, appears in \cite{ishikawa2006}. 
The remaining relations appear to be new. The utility of these relations is that, given the active $\mf{V}$ and $\bm{\Omega}$, they determine the minimal external flow $\mf{v}^{\mrm{act}}(\mf{r})$.  
This $\mf{v}^{\mrm{act}}(\mf{r})$ is the sum of a potential dipole of strength 
$
\mf{d} = - 30 \pi \eta a^3 \left[ 
\la \, (\mf{v}^{\mrm{a}} \cdot \widehat{\mf{r}}) \, \widehat{\mf{r}} \ra + \frac13 \mf{V} 
\right]
$ 
and a spinlet of strength 
$\bm{\Lambda}
   =
   14 \pi \eta a^4
   [
   \langle
   \overbrace{\, \big(\mf{v^{\mrm{a}}} \times \widehat{\mf{r}} \big) \, \widehat{\mf{r}} \, \widehat{\mf{r}} \,}
   \rangle
   - \frac{2a}{5}
   \overbrace{\bm{\Omega} \, \mathbb{I} \, }
   ].
$ 
The stresslet $\mf{S}$, the septlet $\bm{\Gamma}$ and the vortlet $\bm{\Psi}$ modify the external flow without affecting $\mf{V}$ and $\bm{\Omega}$. However, they contribute to long range flows and, thus, influence interparticle hydrodynamic interactions. Eq.\ (\ref{eq:single_sphere_Laddyzhenskaya_relations}) provides a manifestly rotational invariant relationship between the external flow and the rigid body motion, $\mf{V}$ and $\bm{\Omega}$, and active surface velocity $\mf{v}^{\mrm{a}}$ of the particle. Previous work only considered the relationship between rigid body motion and surface velocity confined to purely axisymmetric flows, thus missing the crucial active swirling flow components considered here.

Using Eq.\ (\ref{eq:traction_jump_irred_expansion}) and the linear relation between the stress and velocity multipoles, the power dissipated into the fluid, $ \dot{W} = -\int \mf{q} \cdot \mf{v}^S \, \mrm{d}S$, is obtained in terms of the multipole moments of the stress as 
% 
% \begin{align} \label{eq:power_dissipation}
%   \dot{W} 
%   &= 
%   -\sum_{p=0}^{\infty} \mf{Q}^{(p+1)} \odot \bm{\mathcal{G}}^{(p+1, \, p+1)} \odot \mf{Q}^{(p+1)}, 
% \end{align}
%
$\dot{W} 
  = 
  -\sum_{p=0}^{\infty} \mf{Q}^{(p+1)} \odot \bm{\mathcal{G}}^{(p+1, \, p+1)} \odot \mf{Q}^{(p+1)}, 
$
where the matrix $\bm{\mathcal{G}}$ is diagonal in rank and weight (see \cite{SI}). Resolving into irreducible parts gives 
\begin{align} \label{eq:power_dissipated_in_terms_of_force_multipoles}
  \dot W^{}
  &=
  \frac{3}{20 \pi \eta a^3} \mf{S} \odot \mf{S}
  + \frac{3}{10 \pi \eta a^5} \mf{d} \odot \mf{d}
  + \frac{32}{3 \pi \eta a^5} \bm{\Psi} \odot \bm{\Psi}
  \nonumber \\ 
  & \qquad \qquad
  + \frac{6}{7 \pi \eta a^5} \bm{\Gamma} \odot \bm{\Gamma}
  + \frac{675}{16 \pi \eta a^7} \bm{\Lambda} \odot \bm{\Lambda}.
\end{align}

% ====================================================
% FIGURE # : POWER
% ====================================================
\begin{figure}[tbp]
\includegraphics[width=0.45\textwidth]{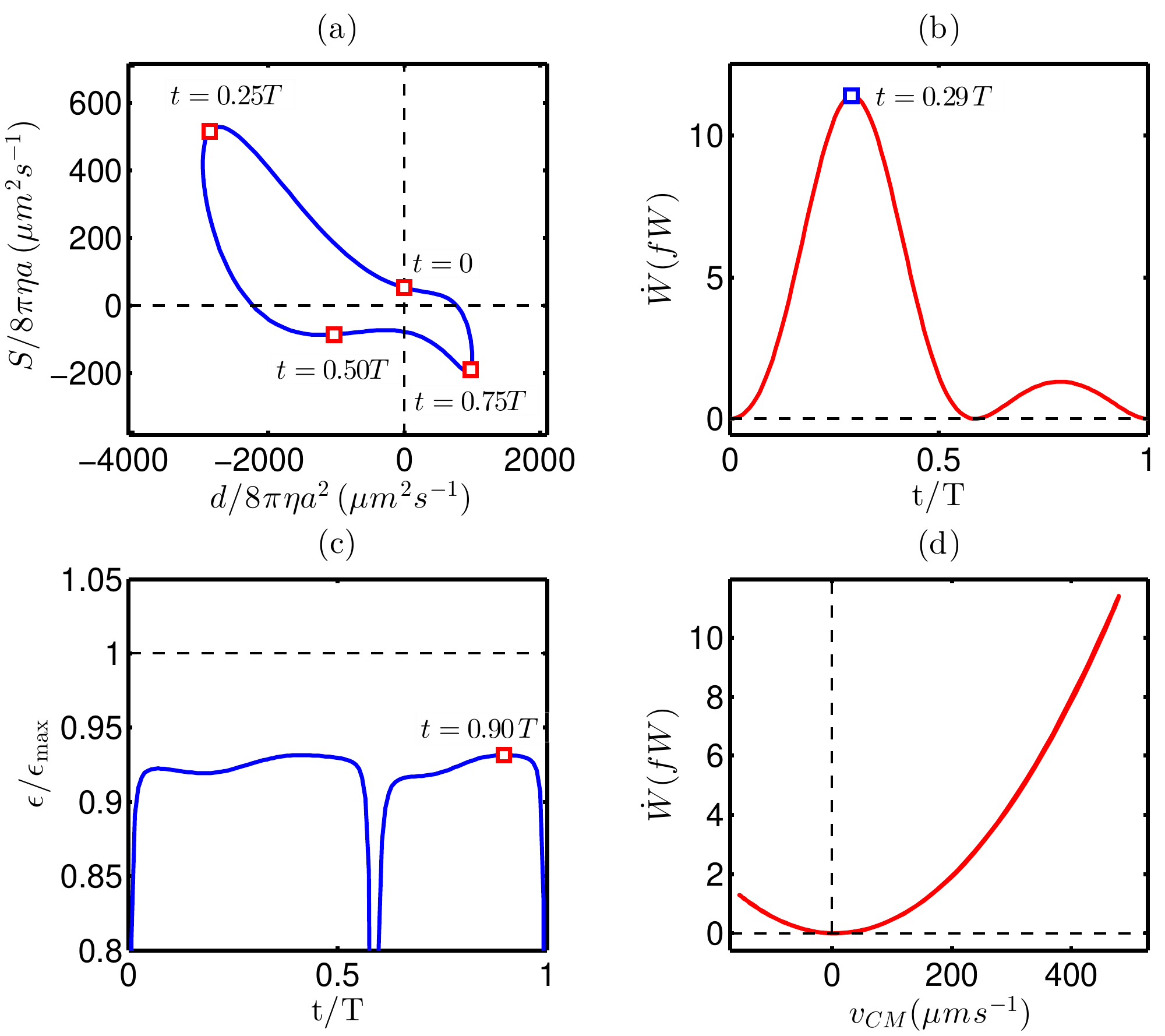}
  \caption{ (Color) Power dissipation and swimming efficiency of Chlamydomonas computed using a simple linear combination of $\mf{S}(t)$, $\mf{d}(t)$ and $\bm{\Gamma}(t)$. (a) Variation of $d_0(t)$ against $S_0(t)$, the former estimated from PIV data of a swimming Chlamydomonas \cite{guasto2010}. (b) Time variation of dissipated power $\dot{W}(t)$. (c) The relative efficiency is maximum near the middle and end of the cycle. (d) Variation of the power $\dot{W}$ against translational velocity $v_{CM}$. 
  }
\end{figure}
% ====================================================

% ====================================================
% SECTION : CHLAMY AND VOLVOX
% ====================================================
\emph{Oscillatory and swirling flows:} The flow around the microorganism \emph{Chlamydomonas reinhardtii} has recently been measured in detail to reveal a flow field that is ``complex and highly time-dependent'' \cite{guasto2010}. We are able to capture the essential features of this flow by superposing flows due to the potential dipole, stresslet, and septlet with time-varying strengths. Assuming particle motion to occur along the $y$-axis the multipoles are parametrized uniaxially as $\mf{S} = S_0(t)(\mf{\widehat{y}} \mf{\widehat{y}} - \frac13 \mathbb{I})$, $\mf{d} = d_0(t) \mf{\widehat{y}}$ and $\bm{\Gamma} = \Gamma_0(t) (\mf{\widehat{y}} \mf{\widehat{y}} \mf{\widehat{y}} - \frac35 \overbrace{\mf{\widehat{y}} \mathbb{I}})$.
The data of \cite{guasto2010} shows that the translation speed can be very well parametrized by the first two Fourier modes, 
$V(t) 
= (a_0/2) \big[
1 + (2 a_1/a_0) \cos(\omega t) + (2 a_2/a_0) \cos(2\omega t) + 
$
$(2 b_1/a_0) \sin(\omega t) + (2 b_2/a_0) \sin(2\omega t) 
\big]$, 
where $a_0$ is in units of $\mu m s^{-1}$ (see \cite{SI}). This yields, through Eq.\ (\ref{eq:single_sphere_Laddyzhenskaya_Force}) and (\ref{eq:single_sphere_Laddyzhenskaya_PotentialDipole}), $d_0(t) = -10 \pi \eta a^3 V(t)$. $S_0(t)$ and $\Gamma_0(t)$ are then determined by the position of the stagnation point relative to the center of the particle. The flow fields produced by this analysis, shown at selected times of the cycle in Fig.\ (\ref{fig:gollub_chlamy}) and in the supplemental video \cite{SI}, are in good agreement with the corresponding figures in \cite{drescher2010, guasto2010}. The particle moves to the right when $d_0 < 0$ and to the left when $d_0 > 0$, with the stagnation point either leading it ($d_0 S_0 < 0$) or lagging behind ($d_0 S_0 > 0$). On average, a Chlamydomonas of size $3.5$ $\mu$m swimming at $134$ $\mu$m$s^{-1}$ in water at $20^{\circ}$C dissipates approximately $6$ fW of power. Both the instantaneous power variation, Fig.\ (2b), and the average power values are in good agreement with experimental findings \cite{guasto2010}. The instantaneous efficiency of translation $\epsilon(t) = 6\pi\eta a V^2/\dot W(t)$ \cite{lighthill1952}, plotted in Fig.\ (2c) has a maximum value close to the theoretical maximum of $20 \%$ (see \cite{SI}). The power dissipation as a function of the speed, shown in Fig.\ (2d), shows the expected quadratic dependence. 

Like most microorganisms, \emph{Volvox carteri} rotates around its own axis as it swims. Using the minimal representation for the spinlet strength, $\Lambda_0 = - (28/5) \pi \eta a^5 \Omega$, and parametrizing uniaxially, $\bm{\Lambda} = \Lambda_0(t) (\mf{\widehat{y}} \mf{\widehat{y}} \mf{\widehat{y}} - \frac35 \overbrace{\mf{\widehat{y}} \mathbb{I}})$, we are able to capture the short-ranged swirling flow field responsible for self-rotation, Fig.\ (S3) \cite{SI}. The vortlet produces swirling flows that spin in opposite directions on the particle surface and thus cancel each other out. Thus the vortlet does not contribute to Volvox rotation (see supplemental Fig.\ (S2) \cite{SI}). Rotation induced by spinlet swirling flows have a maximum swimming efficiency of $1.5 \%$ in the Lighthill sense \cite{lighthill1952}. Representing the Volvox by a uniaxial spinlet whose strength has been computed using its minimal representation, we calculate the rotational power dissipated by a Volvox of size $150 \mu m$ rotating at $1$ rad $s^{-1}$ in water at $20^{\circ} $C to be approximately $250$ fW. 
Swirling flows around Volvox, if experimentally measured, can shed light on the swimming mechanism that must produce the antisymmetric velocity moments on the particle surface.

% ====================================================
% SECTION : STRESS
% ====================================================
\emph{Active stress densities:} 
Stokes flows due to boundary stresses can be reproduced by effective volume force densities $\mf{f}(\mf{r})$ that obey 
$
\int \mf{G}(\mf{r}-\mf{r}') \cdot \mf{f}(\mf{r}') \, \mrm{d}V
= \int \mf{G}(\mf{r}-\mf{r}') \cdot \mf{q}(\mf{r}') \, \mrm{d}S  
$. 
This provides a heuristic for obtaining force densities required for continuum descriptions of active matter.  The active force density that produces the flow in Eq. 4 is 
$\mf{f}(\mf{r}) 
= 
\left\{ \! 1 + (a^2/10) \nabla^2 \! \right\} \! \bm{\nabla} \cdot \mf{S} 
+ \frac15 \nabla^2 \mf{d} 
+ \frac23  \bm{\nabla} \times \left( \bm{\nabla} \cdot \bm{\Psi} \right)  
- \left\{  1 + (a^2/14) \nabla^2  \right\} \bm{\nabla} \bm{\nabla} : \bm{\Gamma} 
- \frac34\bm{\nabla} \times \left( \bm{\nabla} \bm{\nabla} : \bm{\Lambda} \right)
$,
where each of the multipoles are now continuum densities of the form $\mf{S}(\mf{r}) = \sum_n \mf{S}_n \delta(\mf{r} - \mf{r}_n)$. Since the active force density conserves momentum globally, it can always be written as a divergence of an active stress \cite{finlayson1969}, $\mf{f} = \bm{\nabla} \cdot \bm{\sigma}^{\mrm{a}}$, which itself can be decomposed into irreducible symmetric traceless, antisymmetric and isotropic contributions, $\bm{\sigma}^{\mrm{a}} = \bm{\sigma}^s + \frac12 \epsilon \cdot \mf{A} +  \phi \mathbb{I}$. Explicitly these are 
\begin{subequations}
\begin{align}
    \label{eq:stress_tensor_in_terms_of_stress_multipoles_symmetric}
        \bm{\sigma}^s 
    &=
    \left( \! 1 + \frac{a^2}{10} \nabla^2 \! \right) \mf{S}
    - \left( \! 1 + \frac{a^2}{14} \nabla^2 \! \right) 
    \bm{\nabla} \cdot \bm{\Gamma}
    \\
   \label{eq:stress_tensor_in_terms_of_stress_multipoles_antisymmtric}
        \mf{A} 
    &=
    \frac43 \bm{\nabla} \cdot \bm{\Psi}
    - \frac32 \bm{\nabla} \bm{\nabla} : \bm{\Lambda}
    - \frac25 \bm{\nabla} \times \mf{d}
    \\
    \label{eq:stress_tensor_in_terms_of_stress_multipoles_isotropic}
    \phi &= \frac15 \bm{\nabla} \cdot \mf{d}.
\end{align}
\end{subequations}
They enter the balance equation for the momentum density $\mf{g}$,
$\partial_t \mf{g} + \bm{\nabla} \cdot (\mf{gv}) = -\bm{\nabla} p + \eta \nabla^2 \mf{v} + \bm{\nabla} \cdot \bm{\sigma}^s + \frac12 \bm{\nabla} \times \mf{A} + \bm{\nabla} \phi$,  
%\cite{dahler1959}
as additional active sources of momentum transport over the usual pressure and viscous terms.

This heuristic, which can be extended to arbitrary multipolar order, shows that symmetric states of active stress contain contributions beyond the stresslet considered in the literature. Septlet stresses produce $r^{-3}$ flow and are thus important for collective dynamics at long wavelengths. The divergence of the potential dipole density produces an isotropic \emph{active pressure} which has to be balanced by an active flow to ensure incompressibility. 

Equation (\ref{eq:stress_tensor_in_terms_of_stress_multipoles_antisymmtric}) shows that active particles will generically produce antisymmetric states of stress in the fluid. Conservation of orbital angular momentum is violated in the presence of antisymmetric stresses, 
which must be restored by introducing an internal ``spin'' angular momentum $\bm{l}$ such that the total angular momentum, being the sum of orbital and spin contributions, is conserved. Exchanges between orbital and spin contributions, governed by 
$\partial_{t} ( \mathbf{r} \times \mathbf{g} ) + \bm{\nabla} \cdot (\mathbf{r} \times \mathbf{g}\mathbf{v} )  
=
\bm{\nabla} \cdot \big( \mathbf{r} \times \bm{\sigma}^{s} \big ) - \mathbf{A}$, 
$\partial_t \bm{l} + \bm{\nabla} \cdot (\bm{l}\mf{v}) = \bm{\nabla} \cdot \mf{c} + \mf{A}$, 
occur whenever the antisymmetric stresses are non-zero \cite{dahler1959}. Here $\mf{c}$ is the couple stress. Remarkably, and in distinction to antisymmetric stresses in polyatomic  liquids, the antisymmetric active stress $\mf{A}$ has the form of a conserved current, Eq.\  (\ref{eq:stress_tensor_in_terms_of_stress_multipoles_antisymmtric}) and thus separately conserves the global amounts of orbital and spin angular momenta.

Antisymmetric stresses also couple the linear momentum $\mf{g}$ to the angular momenta through $\partial_t \mf{g} + \bm{\nabla} \cdot (\mf{gv}) = -\bm{\nabla} p + \eta \nabla^2 \mf{v} + \bm{\nabla} \cdot \bm{\sigma}^s + \frac12 \bm{\nabla} \times \mf{A} + \bm{\nabla} \phi$  \cite{dahler1959}. This implies that self-rotating particles, through their hydrodynamic interaction, can  set up spontaneous macroscopic flows in suspension. This is a macroscopic manifestation of the translational velocity $(1 + a^2 \nabla^2/6) \,  \mf{v}^{(\Lambda)}$ acquired by a passive particle at $\mf{r}$ due to the flow $\mf{v}^{(\Lambda)}$ produced by a spinlet at the origin. 

% ==============================================
% DISCUSSIONS
% ==============================================
\emph{Discussion:}  The minimal irreducible multipoles introduced here provide an accurate description of the flow, power dissipation and efficiency around active particles. The method, generalized to $N$ particles, allows us to calculate true many-body hydrodynamic interactions between active particles. Applications to suspension rheology beyond the dilute limit, to active flows near rigid boundaries and to the synchronized rigid body motion due to active hydrodynamic flows follow naturally. Our work shows that a second-rank nematic order parameter (corresponding to a density of stresslets) provides only a coarse description of the microstructure of an active suspension, and that a complete description requires order parameters of increasing tensorial rank (corresponding to densities of higher multipoles). We urge the experimental verification of two of our key findings, the swirling flow around a single rotating active particle and the macroscopic advective flows generated in a suspension of rotating particles due to active hydrodynamic interactions. 

Financial support from PRISM II, Department of Atomic Energy, Government of India is gratefully acknowledged. We thank G. Baskaran, M. E. Cates, G. Date, R. E. Goldstein, P. Chaikin, D. Pine, L. Greengard, A. J. C. Ladd, M. Polin, R. Simon, H. A. Stone and P. B. Sunil Kumar for helpful discussions. We thank R. E. Goldstein and M. Polin for kindly sharing their data of \cite{drescher2010} and answering our queries, and A. J. C. Ladd for generously sharing his personal notes as well as for many useful suggestions.

% ====================================================
% SUPPLEMENTAL MATERIAL
% ====================================================
\bigskip
\noindent \textbf{Supplemental material}
\bigskip

% ====================================================
% FIGURE # : STRESSLET, PD, SEPTLET
% ====================================================
\begin{figure*}[tbp]
  \begin{center}
    \subfigure[~Stresslet flow : $\mf{v}^{(\mrm{S})}$]{ 
    \includegraphics[width=0.31\textwidth]{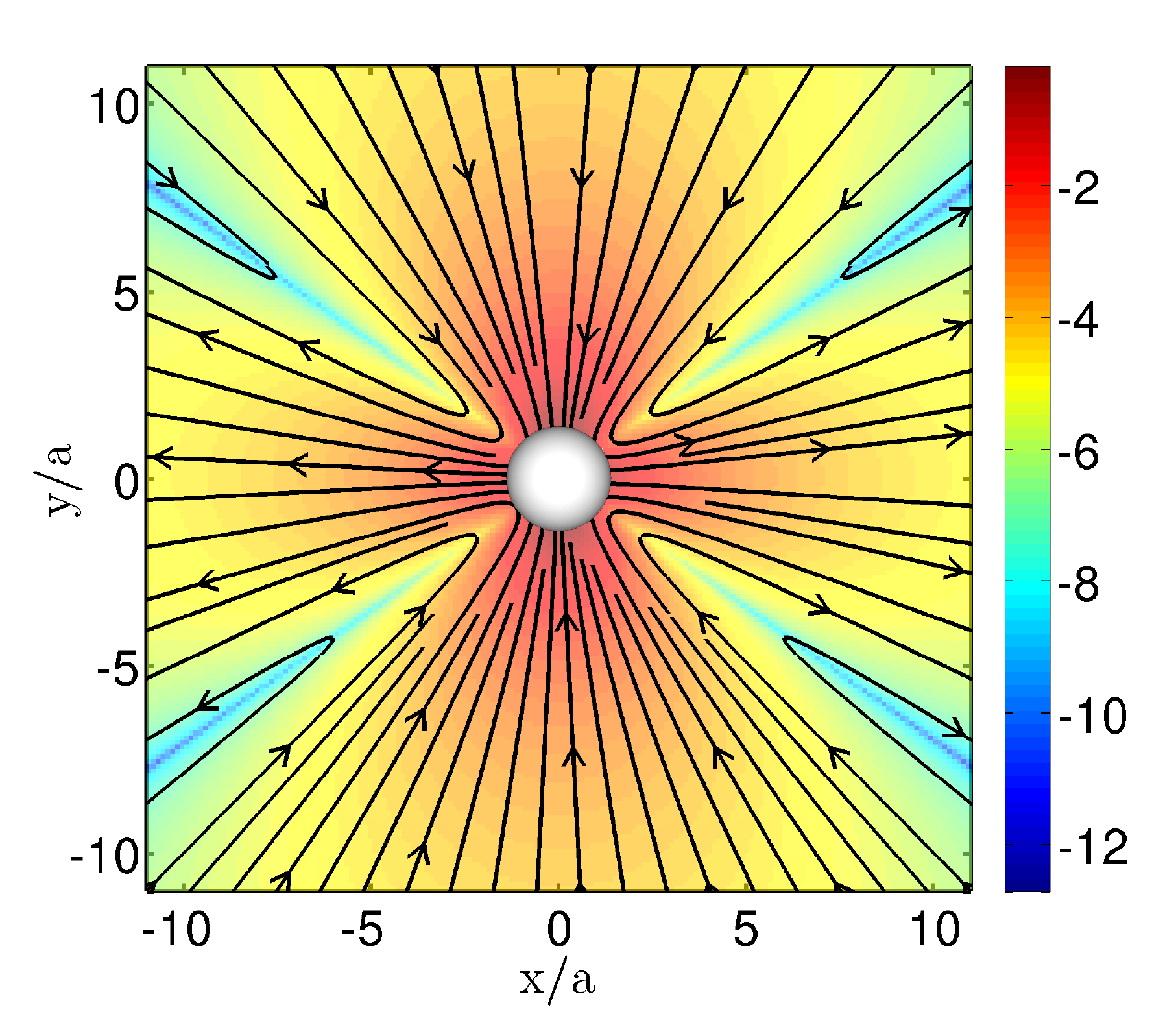} 
    \label{fig:stresslet_color} 
    } %\\
    \subfigure[~Potential dipole flow : $\mf{v}^{(\mrm{d})}$]{ 
    \includegraphics[width=0.31\textwidth]{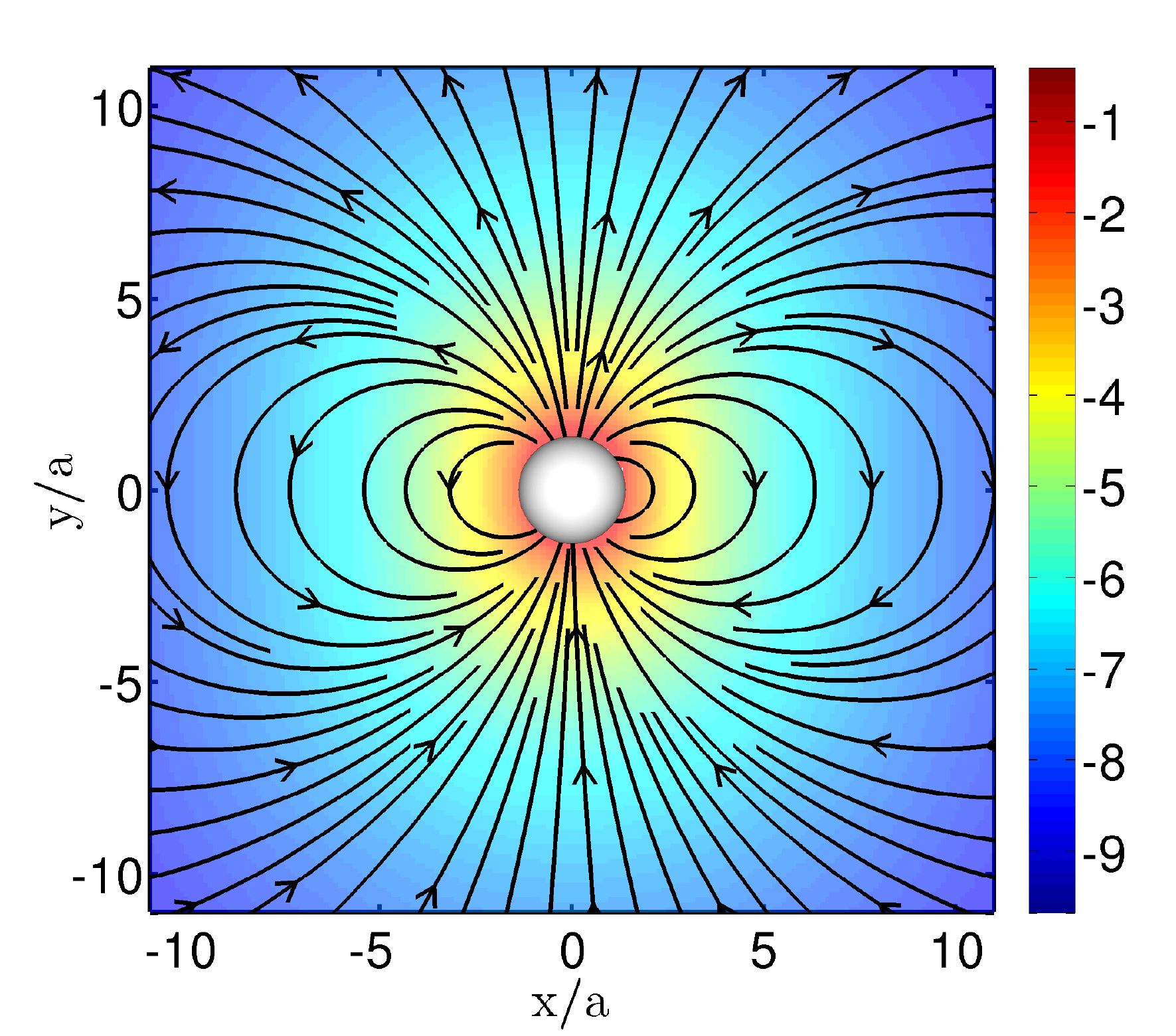} \label{fig:potentialDipole_color} 
    }
    \subfigure[~Septlet flow : $\mf{v}^{(\mrm{\Gamma})}$]{ 
    \includegraphics[width=0.31\textwidth]{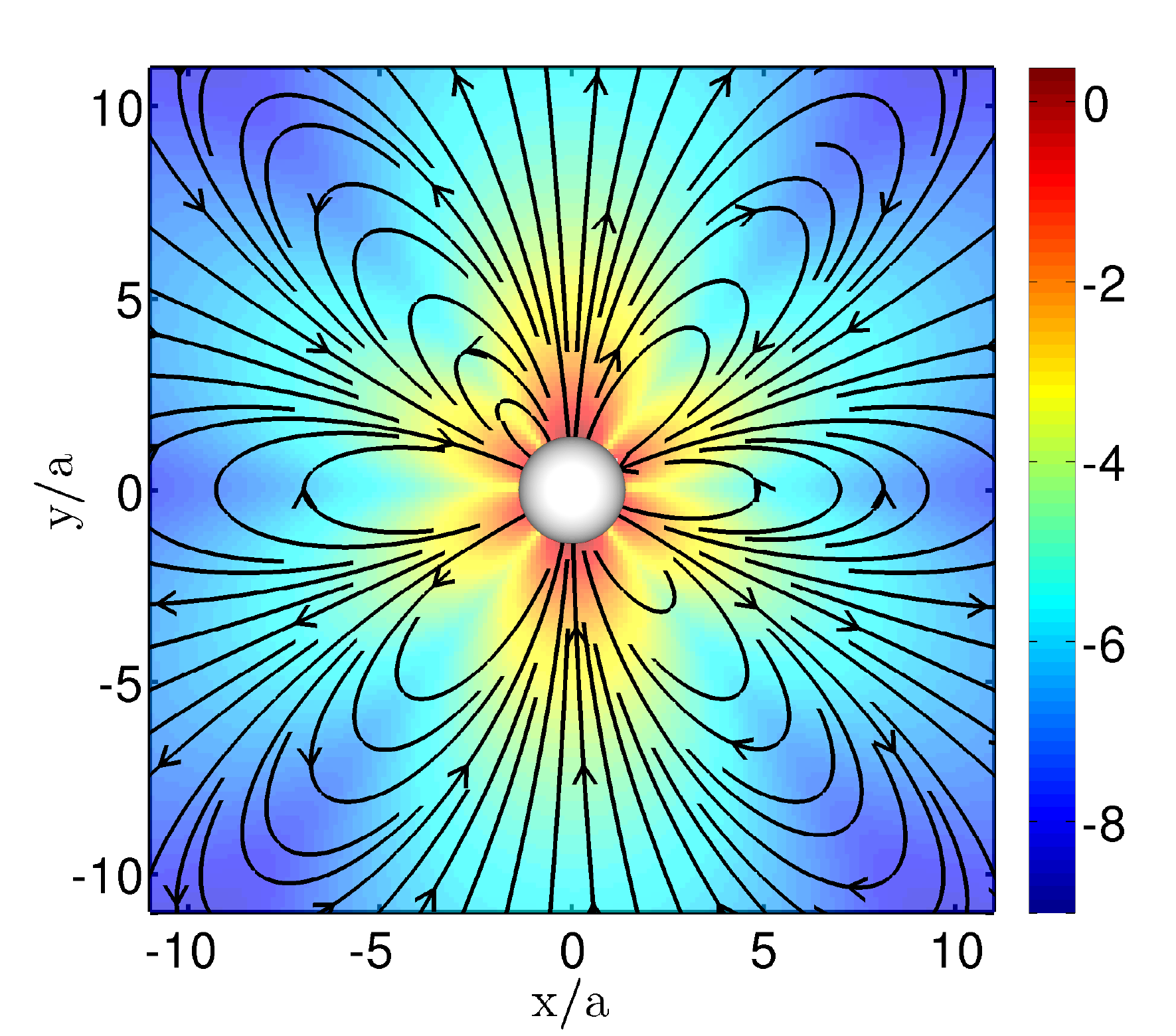} 
    \label{fig:septlet_color} 
    }
  \caption{ FIG.\ S1. Cross-sections of long-range hydrodynamic flows around an active particle of radius $a$ generated by irreducible moments of the single layer density, Eq.\ (4) of main text. Streamlines show the direction of the flow while the background color represents the natural logarithm of the strength of the flow. Panel (a) shows ``puller'' flows generated due to the contractile stresslet $\mf{S}$ uniaxially parametrised in $\unv{y}$. Such flows neither translate nor rotate the particle and decay as $r^{-2}$. The flows due to the potential dipole $\mf{d}$, panel (b), and the uniaxially parametrised septlet $\bm{\Gamma}$, panel (c), generate a net translational effect on the particle along their parametrisation direction $\unv{y}$. These decay as $r^{-3}$.}
  \end{center}
\end{figure*}
% ====================================================

% ====================================================
% FIGURE # : VORTLET
% ====================================================
\begin{figure}[tbp]
  \begin{center}
    \subfigure[~Vortlet flow : $\mf{v}^{(\mrm{\Psi})}$]{ 
    \includegraphics[width=0.45\textwidth]{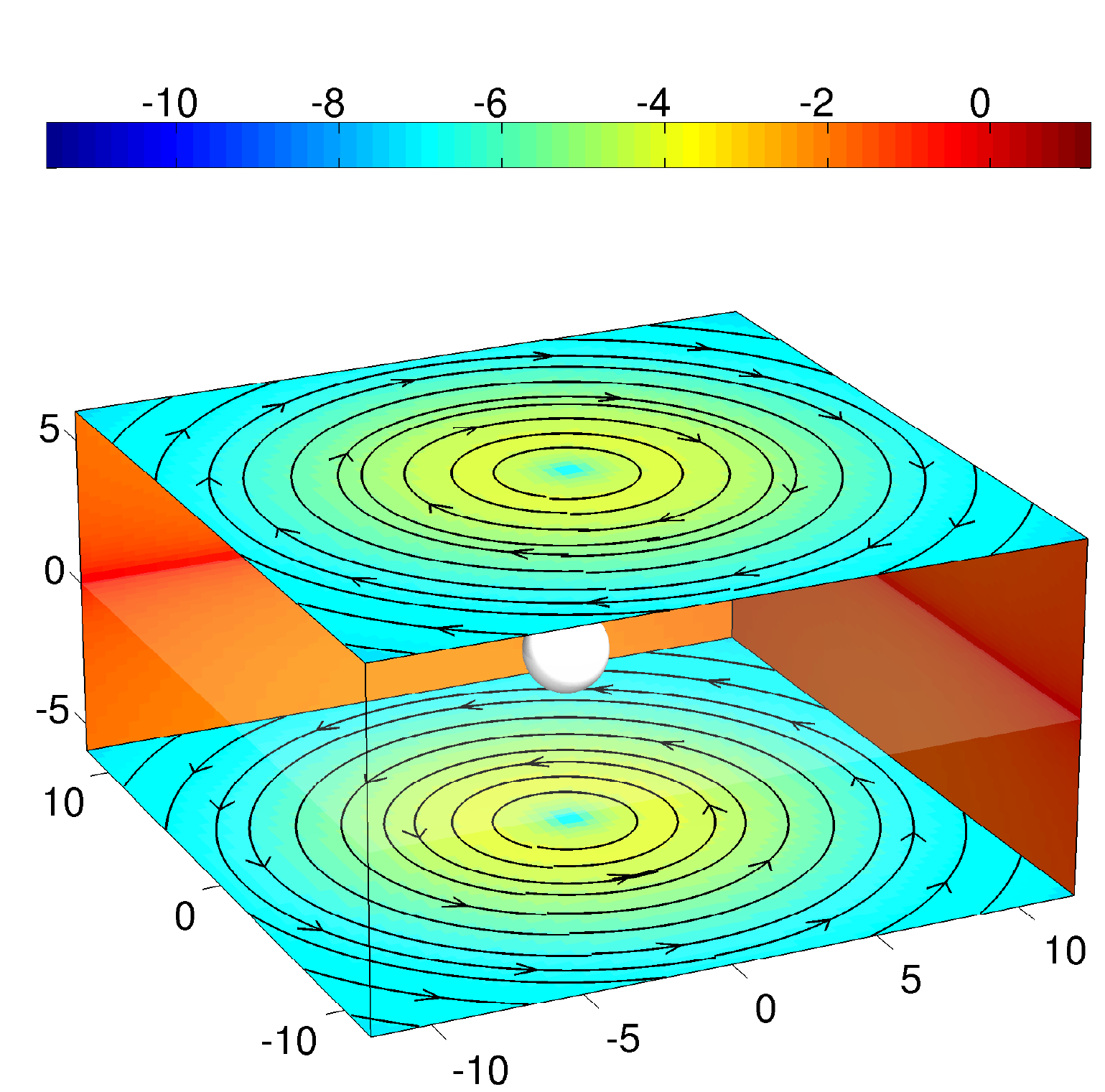} 
    \label{fig:vortlet_color} 
    } 
    \\
    \subfigure[~Spinlet flow : $\mf{v}^{(\mrm{\Lambda})}$]{ 
    \includegraphics[width=0.45\textwidth]{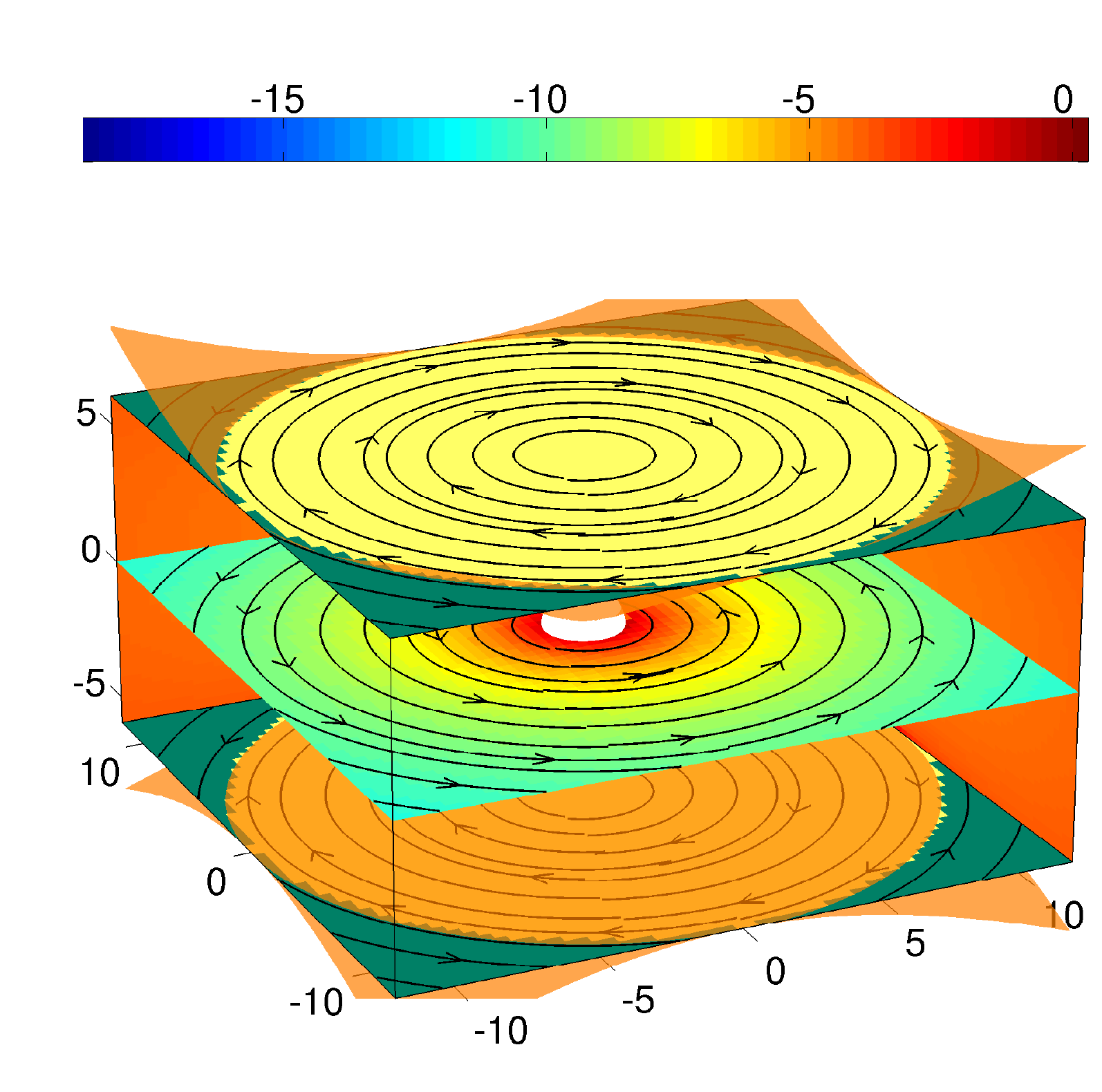} 
    \label{fig:spinlet_color} 
    }
  \caption{ FIG.\ S2.
  Swirling flows around an active particle of radius $a$ generated by uniaxially parametrised irreducible multipole moments of the single layer density, Eq.\ (11) of the main text. Streamlines show the direction of the flow while the background color represents the natural logarithm of the strength of the flow. The vortlet is a dipole of rotlets and thus  produces $r^{-3}$ flows that rotate in opposite directions above and below the equatorial plane $\cos(\theta) = 0$, panel (a), the flows cancelling out on the isosurface. The spinlet produces $r^{-4}$ flows that rotate in the same direction at the particle surface, panel (a), but switch directions across the isosurface $\cos^2(\theta) - 1/5$, where $\theta$ is the polar angle. We predict this to represent swirling flows around Volvox.
  }
  \end{center}
\end{figure}
% ====================================================

% ====================================================
% FIGURE # : CHLAMY FIT
% ====================================================
\begin{figure}[tbp]
  \begin{center} 
    \includegraphics[width=0.45\textwidth]{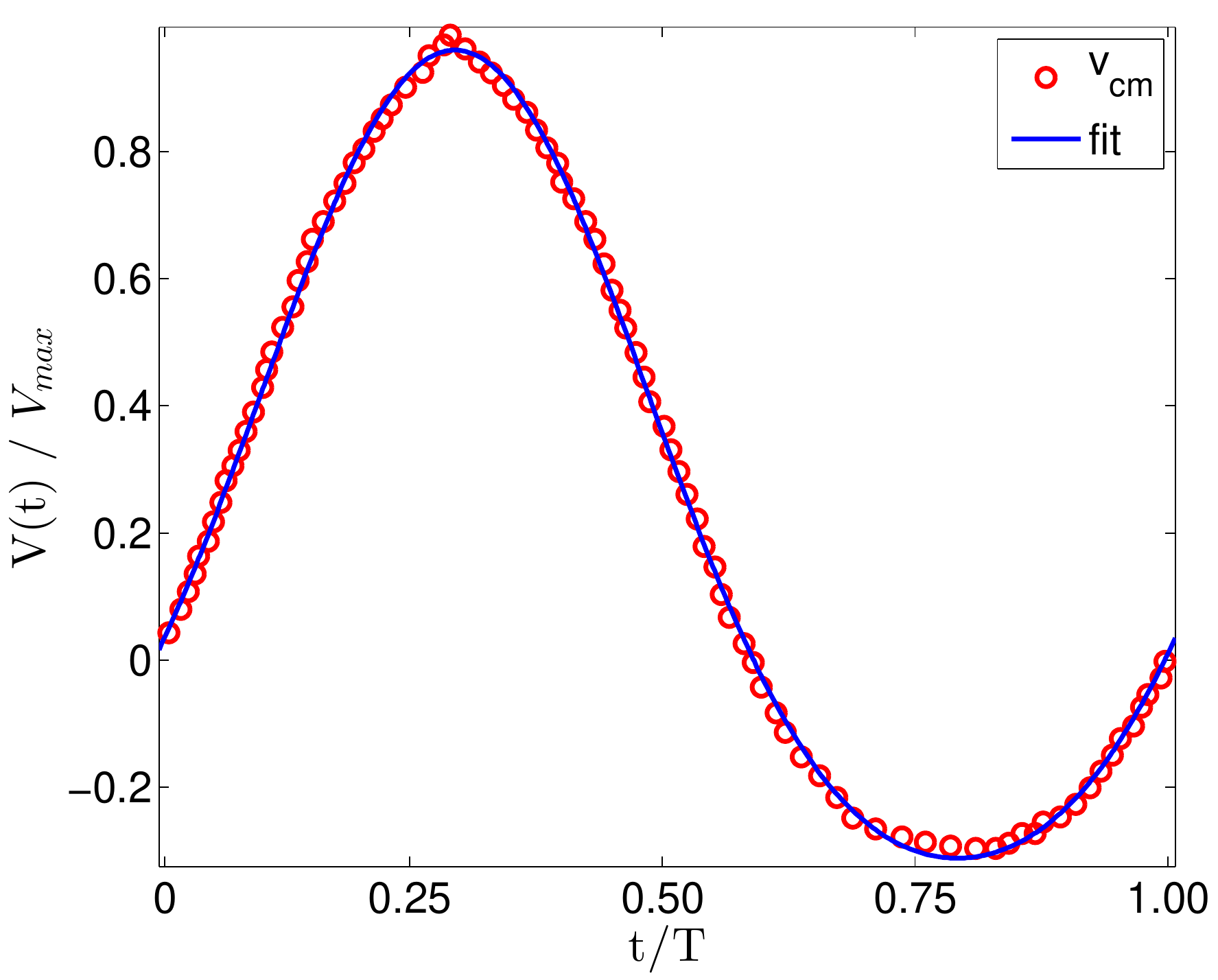} 
  \caption{ FIG.\ S3. Estimation of the centre of mass velocity of Chlamydomonas from flow speeds measured using particle image velocimetry \cite{guasto2010}. Red circles are data values, while the blue line is the Fourier fit, Eq.\ (\ref{eq:chlamy_fourier_fit}), with coefficients as given in the text.
  \label{fig:chlamy_fit} 
  }
  \end{center}
\end{figure}
% % ===============================================================================================%

\newpage
% =====================================================================================
% SECTION
% =====================================================================================
\emph{Solution of Stokes flow using irreducible expansions} : The Stokes equation for chemomechanically active flows is reformulated using boundary integrals \cite{ladyzhenskaya1969, pozrikidis1992, kim2005}, 
\begin{align} \label{eq:boundary_integral_representation_and_equation}
  \int_{S'}
  \mf{G}(\mf{r} - \mf{r}') \cdot \mf{q}(\mf{r}') \, \mrm{d} S'
  &=
  - 8 \pi \eta \, 
  \begin{cases}
    \mf{v}(\mf{r}), \quad \mf{r} \in V
    \\
    \mf{v}^S(\mf{r}), \quad \mf{r} \in S
  \end{cases}
\end{align}
where $\mf{v}(\mf{r})$ is the flow velocity in the bulk while $\mf{v}^S(\mf{r})$ is that on the boundary, that is, the surface of the particle. The Green's function $\mf{G}(\mf{r}) = (\mathbb{I} + \unv{r} \unv{r})/|\mf{r}|$ propagates the single layer density $\mf{q}(\mf{r})$ on the particle surface. If surface stresses $\bm{\sigma}^S$ are known, then the Neumann problem can be solved for spherical boundaries by expanding the single layer density in a spherical harmonic basis. A manifestly rotational covariant representation of such a basis is provided by irreducible Cartesian tensors $\ovb{r}{p}$, which obey the orthogonality condition
\begin{align} \label{eq:irred_orthogonality}
  \left\la \ovb{r}{p} \ovb{r}{q} \right\ra 
  &= 
  \frac{p!}{(2p+1)!!} \, \delta_{p,q} \, \bm{\Delta}^{(p,p)}. 
\end{align}
The surface average $\la \ldots \ra = (1/4 \pi a^2) \int \mrm{d} S$, while the rank $2p$ tensor $\bm{\Delta}^{(p,p)}$ projects any $p$-th rank tensor $\mf{\widehat{r}}^{(p)}$ to its irreducible form $\ovb{r}{p}$ \cite{hess1980, mazur1982}. 
Expanding $\mf{q}(\mf{r})$ in this basis, we get
\begin{align} %\label{eq:traction_jump_irred_expansion}
  \mf{q}(\mf{r}) 
  &= 
  \sum_{p=0}^{\infty}
  \frac{( 2p + 1 ) !!}{4 \pi a^2}
  \, \ovb{r}{p} \odot \, \mf{Q}^{(p+1)}, \quad \mf{r} \in S,
\end{align}
where the multipole moments $\displaystyle {Q}^{(p+1)}_{i\alpha_1 \ldots \alpha_p}$, symmetric and traceless in the last $p$ indices, are given by
\begin{align}
  \mf{Q}^{(p+1)}
  &=
  \frac{1}{p!} 
  \int_S \mf{q}(\mf{r}) \ovb{r}{p} \mrm{d} S.
\end{align}
In Eq.\ (\ref{eq:traction_jump_irred_expansion}), the symbol $\odot$ represents a $p$-fold contraction between a $p$-th rank tensor and another of higher rank, contracting the last index of the first tensor with the first index of the latter till $p$ indices are contracted, such that $\ovb{r}{p} \odot \, \mf{Q}^{(p+1)} = \overbracket[0.7pt]{\widehat{r}_{\alpha_1 \alpha_2 \ldots \alpha_{p-1} \alpha_p}} {Q}^{(p+1)}_{\alpha_p \alpha_{p-1} \ldots \alpha_2 \alpha_1 \, i}$. 
We now insert Eq.\ (\ref{eq:traction_jump_irred_expansion}) into Eq.\ (\ref{eq:boundary_integral_representation_and_equation}), $\mf{r} \in V$, and obtain
\begin{align}
  8 \pi \eta \, \mf{v}(\mf{r})
  &= 
  - \sum_{p=0}^{\infty}
  \frac{( 2p + 1 ) !!}{4 \pi a^2}
  \int_{S'} \mf{G}(\mf{r}-\mf{r}') \cdot 
  \mf{Q}^{(p+1)} \, \odot \ovb{r}{p} \, 
  \mrm{d} S'
\end{align}
Writing this in terms of the Fourier transform of the Green's function, $\mf{G}(\mf{k}) = 8 \pi (\mathbb{I} - \unv{k} \, \unv{k} ) / |\mf{k}|^2$, we get
\begin{align} \label{eq:bulk_flow_in_fourier_before_plane_wave_expansion}
  8 \pi \eta \, \mf{v}(\mf{r}) 
  &=
  - \sum_{p=0}^{\infty}
  \frac{( 2p + 1 ) !!}{4 \pi a^2}
  \int_k
  \frac{\mrm{d}^3 k}{(2\pi)^3} \, 
  e^{i \mf{k} \cdot \mf{r} } \, 
  \mf{G}(\mf{k}) \cdot \mf{Q}^{(p+1)} \, 
  \nonumber \\
  & \qquad \qquad 
  \odot 
  \int_{S'}
  \mrm{d} S'
  e^{-i \mf{k} \cdot \mf{r}' } \ovb{r}{p} \, 
\end{align}
We expand the plane wave in spherical Bessel functions,
\begin{align} \label{eq:spherical_wave_expansion}
  e^{i \mf{k} \cdot \mf{r}} 
  &= 
  \sum_{m = 0}^{\infty} 
  \frac{(i)^m (2m+1)!!}{m!} 
  j_m(kr) 
  \ovb{k}{m}
  \odot 
  \ovb{r}{m},
\end{align}
and thus obtain
\begin{widetext}
\begin{align} \label{eq:bulk_flow_in_fourier_after_plane_wave_expansion}
  8 \pi \eta \, \mf{v}(\mf{r}) 
  &=
  - \sum_{p=0}^{\infty} \sum_{m=0}^{\infty}
  \frac{(-i)^m (2p+1)!! (2m+1)!!}{m!}
%   \nonumber \\
%   & \qquad \times
  \int \frac{\mrm{d}^3 k}{(2 \pi)^3} \,
  e^{i \mf{k} \cdot \mf{r}} j_m(k a) \,
  \mf{G}(\mf{k}) \cdot \mf{Q}^{(p+1)} 
%   \nonumber \\
%   & \qquad \qquad 
  \odot
  \left\la
  \ovb{r}{p} \ovb{r}{m}
  \right\ra
  \odot \ovb{k}{m}
  \nonumber \\
  &=
  - \sum_{p=0}^{\infty} 
  (-i)^p (2p+1)!! 
%   \nonumber \\
%   & \qquad \times
  \int \frac{\mrm{d}^3 k}{(2 \pi)^3} \,
  e^{i \mf{k} \cdot \mf{r}} j_p(k a) \,
  \mf{G}(\mf{k}) \cdot \mf{Q}^{(p+1)} \odot
  \unv{k}^{(p)}.
\end{align}
\end{widetext}
Here $\bm{\Delta}^{(p,p)}  \odot \ovb{k}{p} = \ovb{k}{p}$, and $\mf{Q}^{(p+1)} \odot \ovb{k}{p} \equiv \mf{Q}^{(p+1)} \odot \, \unv{k}^{(p)}$ since $\mf{Q}^{(p+1)}$ is already symmetrized and detraced in its trailing $p-1$ indices. The spherical Bessel function can be expanded in polynomials of the wavenumber $k$ and truncated, 
\begin{align}
  j_p(ka)
  &=
  \frac{a^p k^p}{(2p+1)!!} 
  \left[ 
  1 - \frac{a^2 k^2}{4p+6} + \mathcal{O}(k^4)
  \right]
  \nonumber \\
  &
  =
  \frac{a^p k^p}{(2p+1)!!} 
  \left(  1 - \frac{a^2 k^2}{4p+6}  \right)
\end{align}
since biharmonicity ensures that $\mathcal{F} \left\{ k^4 \mf{G}(\mf{k}) \right\} = 0$, where $\mathcal{F}$ is the Fourier transform operator. Substituting this in Eq.\ (\ref{eq:bulk_flow_in_fourier_after_plane_wave_expansion}), we get the required flow equation
% 
% \begin{widetext}
\begin{align} %\label{eq:stokes_flow_for_single_sphere}
  & 8 \pi \eta \, \mf{v}(\mf{r})
  =
  - \sum_{p=0}^{\infty}
  a^p \, \mf{Q}^{(p+1)} 
  \nonumber \\
  & \qquad \odot
  \int \frac{\mrm{d}^3 k}{(2 \pi)^3} \, e^{i \mf{k} \cdot \mf{r}}
  \left( -i \mf{k} \right)^{(p)}\cdot
  \left(  1 - \frac{a^2 k^2}{4p+6}  \right) \mf{G}(\mf{k})
  \nonumber \\
  & \qquad 
  =
  - \sum_{p=0}^{\infty}
  a^p \, \mf{Q}^{(p+1)} \odot
  (-\bm{\nabla})^{(p)} \cdot
  \left(  1 + \frac{a^2}{4p+6} \nabla^2 \right) \mf{G}(\mf{r})
\end{align}
% \end{widetext}

% =====================================================================================
% SECTION
% =====================================================================================
\emph{Resolving surface stress multipoles into irreducible parts} : The general solution in Eq. (\ref{eq:stokes_flow_for_single_sphere}) can be simplified by decomposing the reducible single layer moments into irreducible tensors. Any $p$-th rank tensor $\mf{Q}^{(p)}$ can be decomposed into irreducible tensors $\mf{Q}^{(p \, ; \, j, r)}$ of weight $j \leq p$ with $2j+1$ independent components, subtending a $j$ dimensional irreducible representation of the rotation group $SO(3)$ \cite{coope1965, jerphagnon1970, jerphagnon1978}. The seniority index $r$ is needed when more than one weight $j$ representation occurs in the decomposition. The general decomposition is then the direct sum $\mf{Q}^{(p)} = \oplus_{j,r} \mf{Q}^{(p \, ; \, j, r)}$ \cite{coope1965, jerphagnon1970, jerphagnon1978}. The tensors corresponding to weights $1, 2$ and $3$ are known as vectors, deviators, and septors respectively and can be further classified by their parity as polar (true) or axial (psuedo) tensors. Here we focus on the minimal set of multipoles required to produce active translations and rotations. This requires us to enumerate all irreducible multipoles  $p \leq 2$ and the pseudoseptorial multipole for $p=3$. The decompositions we require are given in tensorial form in the main text. Here we present the same in index notation,
\begin{subequations} \label{eq:traction_multipoles_index_form}
  \begin{align}
  Q_i^{(1)} 
  &= 
  F_i, 
\\
  Q_{i \alpha}^{(2)} 
  &=  
  \frac{1}{a} \left[ S_{i \alpha} - \frac12 \epsilon_{i \alpha \nu} T_{\nu} \right], 
\\
  Q_{i \alpha \beta}^{(3)} 
  &=
  \frac{1}{a^2}
  \left[
  \Gamma_{i \alpha \beta} 
  + \frac23 \left( \epsilon_{i \alpha \nu} \Psi_{\nu \beta} + \epsilon_{i \beta \nu} \Psi_{\nu \alpha }\right)
  \right.
  \nonumber \\
  & 
  \left.
  + \frac{1}{10} \left(-2 d_i \delta_{\alpha \beta} + 3 d_{\alpha} \delta_{\beta i} + 3 d_{\beta} \delta_{i \alpha} \right) 
  \right]
\\
  Q^{(4; \, 3)}_{i \, \alpha \beta \gamma} 
  &= 
  - \frac{1}{4 a^3} \left(\epsilon_{i \alpha \nu} \Lambda_{\nu \beta \gamma} + \epsilon_{i \beta \nu} \Lambda_{\nu \gamma \alpha} + \epsilon_{i \gamma \nu} \Lambda_{\nu \beta \alpha} \right).     
\end{align}
\end{subequations}
where $\bm{\epsilon}$ is the rank-3 antisymmetric Levi-Civita tensor. As stated in the main text, the new irreducible multipoles introduced here are the pseudodeviatoric torque dipole $\bm{\Psi}$ or the ``vortlet'', the septorial stresslet dipole $\bm{\Gamma}$ or the ``septlet'', and the pseudoseptorial multipole $\bm{\Lambda}$ or the ``spinlet''.

% =====================================================================================
% SECTION
% =====================================================================================
\emph{Uniaxial parametrisations of stress multipoles} : Uniaxial parametrisations are the simplest representations of the stress multipoles. Let $\unv{p}$ determine the parametrisation direction. The vectorial potential dipole of strength $d_0$ is then trivially parametrised as $\mf{d} = d_0 \unv{p}$. The deviatoric $\mf{S}$ is parametrised as $S_0 (\unv{p} \unv{p} - \frac13 \mathbb{I})$, while the pseudodeviatoric $\bm{\Psi}$ takes a similar form $\Psi_0 (\unv{p} \unv{p} - \frac13 \mathbb{I})$. The septorial $\bm{\Gamma}$ is parametrised as $\Gamma_0 (\unv{p} \unv{p} \unv{p} - \frac35 \overbrace{\unv{p} \mathbb{I}})$, while the pseudoseptorial $\bm{\Lambda}$ takes a similar form $\Lambda_0(\unv{p} \unv{p} \unv{p} - \frac35 \overbrace{\unv{p} \mathbb{I}})$. The stresslet strength $S_0$ and the septlet strength $\Gamma_0$ are true scalars, while the vortlet strength $\Psi_0$ and the spinlet strength $\Lambda_0$ are  pseudoscalars. These parametrisations preserve the symmetry and tracelessness conditions of the polar and axial deviators and septors.

% =====================================================================================
% SECTION
% =====================================================================================
\emph{Relation between surface stress and velocity multipoles} : If the velocities on the boundaries are known, then the resulting Dirichlet problem is solved by expanding the prescribed the surface velocity $\mf{v}^S(\mf{r})$ in the same irreducible basis $\ovb{r}{p}$,
\begin{align} \label{eq:surface_velocity_irred_expansion}
    \mf{v}^S (\mf{r})
    &= 
    \sum_{p=0}^{\infty} 
    \frac{1}{p!} \, \ovb{r}{p} \odot \, \mf{V}^{(p+1)}
\end{align} 
where the multipole moments $\displaystyle {V}^{(p+1)}_{i\alpha_1 \ldots \alpha_p}$, symmetric and traceless in the last $p$ indices, are given by 
\begin{align}
 \mf{V}^{(p+1)} &= 
 (2p+1)!! \big\la \mf{v}^{\mrm{S}} \ovb{r}{p} \, \big\ra.
\end{align}
Equating Eq.\ (\ref{eq:surface_velocity_irred_expansion}) and Eq.\ (\ref{eq:bulk_flow_in_fourier_after_plane_wave_expansion}) on $S$, and expanding the plane wave in spherical Bessel functions once again, we get
\begin{align}
  & 8 \pi \eta \, 
  \sum_{m=0}^{\infty} \frac{1}{m!} \ovb{r}{m} \odot \mf{V}^{(m+1)}
  =
  \nonumber \\
  & \qquad 
  - \sum_{p=0}^{\infty} \sum_{n=0}^{\infty} 
  \frac{1}{n!} i^{n-p} (2p+1)!! (2n+1)!!
  \nonumber \\
  & \times 
  \int \frac{d^3 k}{(2 \pi)^3}
  j_n(ka) j_p(ka)
  \ovb{k}{n} \odot \ovb{r}{n}
  \mf{G}(\mf{k}) \cdot \unv{k}^{(p)}
  \odot \mf{Q}^{(p+1)} 
\end{align}
Using the spherical Bessel function identity 
\begin{align}
  \int_0^{\infty} \frac{dk}{2\pi} j_n(ka) j_p(ka) 
  &= 
  \frac{1}{4 a (2n+1) }\delta_{n,p} 
\end{align}
and orthogonality, Eq.\ (\ref{eq:irred_orthogonality}), we get 
\begin{align}
  & 8 \pi \eta \, 
  \frac{1}{(2p + 1)!!} \mf{V}^{(p+1)}
  =
  \nonumber \\
  & \qquad
  - \frac{(2p-1)!!}{4 \pi a}
  \int \frac{k^2 \mrm{d} \Omega_k}{4 \pi} \,
  \unv{k}^{(p)} \, \mf{G}(\mf{k}) \cdot \unv{k}^{(p)}
  \odot \, \mf{Q}^{(p+1)} 
\end{align}
Since $k^2\mf{G}(\mf{k}) = 8 \pi (\mathbb{I} - \unv{k} \unv{k})$, we finally get the desired relation
\begin{align} \label{eq:single_sphere_V_F_relation}
  \mf{V}^{(p+1)} 
  &=
  \bm{\mathcal{G}}^{(p+1,\,p+1)} \odot
  \mf{Q}^{(p+1)} 
\end{align}
where the $2(p + 1)$ rank tensor $\bm{\mathcal{G}}$ is given by \cite{ladd1988}
\begin{align} \label{eq:single_sphere_connector_matrix_}
  \bm{\mathcal{G}}^{(p+1, \, p+1)} 
  &= 
  - 
  \frac{(2p-1)!! (2p+1)!!}{ \left( 4 \pi \eta a \right) }
  \int \frac{d \Omega}{4 \pi} \, 
  \overbracket[0.7pt]{ \, \widehat{\mf{r}}^{(p)} }
  \left( \mathbb{I} - \unv{r} \unv{r} \right) 
  \overbracket[0.7pt]{ \, \widehat{\mf{r}}^{(p)} }
\end{align}

% =====================================================================================
% SECTION
% =====================================================================================
\emph{Estimation of Chlamydomonas and Volvox flows} : We estimate the flow speed data in \cite{guasto2010} by the first two Fourier modes,
\begin{align} \label{eq:chlamy_fourier_fit}
  V(t) 
  &= 
  \frac12 a_0 + a_1 \cos(\omega t) + a_2\cos(2\omega t) 
  \nonumber \\
  & \qquad
  + b_1 \sin(\omega t) + b_2 \sin(2\omega t)
\end{align}
where the values are given by $a_0 = 247.7$, $a_1 = -86.81$, $a_2 = -31.9$, $b_1 = 305.6$ and $b_2 = -21.1$, all in units of $\mu m s^{-1}$. We show the values and the corresponding fit in Fig.\ (S3). Since active flows are force-free, we know from Eq.\ (5a) of the  main text  that $\big\la v^{\mrm{a}}(t) \big\ra = - V(t)$. Using Eq.\ (5d) of the main text we now extract the strength of the potential dipole using the minimal representation, $d_0(t) = -10 \pi \eta a^3 \big\la v^{\mrm{a}}(t) \big\ra$, and we estimate the stresslet and the septlet strengths from the position of the stagnation point. Using Eq.\ (4) of the main text, we linearly combine the flows due to $\mf{v}^{(\mrm{S})}$, $\mf{v}^{(\mrm{d})}$ and $\mf{v}^{(\mrm{\Gamma})}$. The result, shown in the supplementary video and Fig.\ (1) of the main text, effectively captures essential features of the flow around a swimming Chlamydomonas \cite{guasto2010}. From Eq.\ (7) of the main text the power dissipated by the Chlamydomonas is
\begin{align}
\dot{W}^{\mrm{Ch}}(t)
  =
  \frac{3}{20 \pi \eta a^3} \mf{S} \odot \mf{S}
  + \frac{3}{10 \pi \eta a^5} \mf{d} \odot \mf{d}
  + \frac{6}{7 \pi \eta a^5} \bm{\Gamma} \odot \bm{\Gamma}
\end{align}
The instantaneous efficiency of translation, defined as ratio of power expended by an external force to maintain a rigid sphere in uniform motion with speed $V$ to that expended chemomechanically to maintain the same speed \cite{lighthill1952}, is computed to be 
\begin{align}
 \epsilon^{\mrm{Ch}}(t) 
 &= 
 \frac{6\pi\eta a V^2}{\dot{W}^{\mrm{Ch}}(t) }
\end{align}
We have, using Eq.\ (5d) of the main text, $\mf{d} \odot \mf{d} = 100 \pi^2 \eta^2 a^6 V^2$ for purely tangential surface flows. Therefore the maximum translational efficiency is $20\%$.

Near-field swirling flows around Volvox are obtained using the uniaxially parametrised spinlet multipole. Although the vortlet too generates swirling flows, it does not give rise to particle rotations since the flows spin in opposite directions above and below the equatorial plane of the particle and thus cancel out, Fig.\ (S2). From Eq.\ (7) of the main text the power dissipated by the Volvox is
\begin{align}
  \dot{W}^{\mrm{Vo}}(t)
  =
  \frac{675}{16 \pi \eta a^7} \bm{\Lambda} \odot \bm{\Lambda},
\end{align}
Extending Lighthill's definition \cite{lighthill1952}, we define the rotational efficiency as 
\begin{align}
 \epsilon^{\mrm{Vo}}(t) 
 &= 
 \frac{8 \pi\eta a^3 \Omega^2}{\dot{W}^{\mrm{Vo}}(t) }.
\end{align}
Setting terms like $\langle (\mf{v}^{\mrm{a}} \times \widehat{\mf{n}}) \widehat{\mf{n}} \widehat{\mf{n}} \rangle$ to zero, we get $\bm{\Lambda} \odot \bm{\Lambda} = \frac{1568}{125} \pi^2 \eta^2 a^{10} \Omega^2$. The maximum rotational efficiency thus comes out be approximately $1.5 \%$.

% % ====================================================
% % % BIBLIOGRAPHY (OUTPUT OF .BBL)
% % ====================================================
%
% % ====================================================

% ================== Th-th-that's all, folks! ==================================
\end{document}